\documentclass[12pt]{iopart}
\usepackage{citesort}
\usepackage[left=2cm,top=2cm,bottom=2cm,right=2cm]{geometry}
\usepackage{graphicx,enumerate,bm,color}
\expandafter\let\csname equation*\endcsname\relax
\expandafter\let\csname endequation*\endcsname\relax
\usepackage{amsmath,amssymb}

\newcommand{\tst}{\textstyle}

\newcommand{\dsp}{\displaystyle}
\newcommand{\mbf}{\mathbf}
\newcommand{\mrm}{\mathrm}
\newcommand{\ud}{\mathrm{d}}
\newcommand{\be}{\begin{equation}}
\newcommand{\ee}{\end{equation}}
\newcommand{\bea}{\begin{eqnarray}}
\newcommand{\eea}{\end{eqnarray}}

\begin{document}
\title[Multichannel quantum-defect theory for ultracold atom-ion collisions]{Multichannel quantum-defect theory for ultracold atom-ion collisions}
\author{Zbigniew Idziaszek}
\address{Faculty of Physics, University of Warsaw,
00-681 Warsaw, Poland}
\author{Andrea Simoni}
\address{Institut de Physique de Rennes, UMR 6251 du CNRS and Universit\'e de
Rennes 1, 35042 Rennes Cedex, France}
\author{Tommaso Calarco}
\address{Institute of Quantum Information
Processing, University of Ulm, D-89069 Ulm, Germany}
\author{Paul S. Julienne}
\address{Joint Quantum Institute, NIST and
the University of Maryland, Gaithersburg, Maryland 20899-8423, USA}

\begin{abstract}
We develop an analytical model for ultracold atom-ion collisions using
the multichannel quantum-defect formalism. The model is based on the
analytical solutions of the $r^{-4}$ long-range potential and on the
application of a frame transformation between asymptotic and molecular
bases. This approach allows the description of the atom-ion interaction
in the ultracold domain in terms of three parameters only: the singlet and
triplet scattering lengths, assumed to be independent of the relative
motion angular momentum, and the lead dispersion coefficient of the
asymptotic potential. We also introduce corrections to the scattering
lengths that improve the accuracy of our quantum-defect model for
higher order partial waves, a particularly important result for an accurate
description of shape and Feshbach resonances at finite temperature. The
theory is applied to the system composed of a ${}^{40}$Ca$^{+}$ ion and
a Na atom, and compared to numerical coupled-channel calculations carried
out using {\it ab initio} potentials. For this particular system, we investigate
the spectrum of bound states, the rate of charge-transfer processes,
and the collision rates in the presence of magnetic Feshbach resonances at
zero and finite temperature.  \end{abstract}


\maketitle

\section{Introduction}

The successful realization of systems combining ultracold atoms and
ions is stimulating an increasing interest in the physics of atom-ion
collisions at very low collision energy. The first experiments
were realized for clouds of atoms and ions stored in hybrid dual
charged-neutral traps at mK temperatures \cite{Smith2005,Grier2009}. More
recent experiments performed with atoms in the micro and nano Kelvin
temperature domain consist in immersing single ions in a Bose-Einstein
condensate \cite{Zipkes2010,Zipkes2010a,Schmid2010}. Understanding
atom-ion collisions in the quantum regime is an essential elementary
step before a more complete many-body description of these systems
can be developed. While atom-ion collision properties are well
known for high collision energies ~\cite{Bransden1992,Delos1981}, a
theoretical description in the ultracold domain is still under development
\cite{Cote2000,Bodo2008,Idziaszek2009,Gao2010}. Atom-ion and neutral atoms
scattering are significantly different, mainly due to the relatively
long-range character of the polarization interaction between atom and
ion. Beyond the fundamental interest of their collision properties, these
systems are also very attractive for quantum information processing
\cite{Idziaszek2007,Doerk2010}. Hybrid architectures may profit from
advantages offered by both ionic and atomic species, namely a short
computation time for charged particles and a long-coherence time for
neutral atoms. Ultracold charged-neutral systems are also expected to
exhibit interesting many-body effects, including for instance nontrivial
modifications of the condensate wave function in the presence of ionic
impurities \cite{Massignan2005} or the creation of mesoscopic size
molecular ions \cite{Cote2002}.


In a previous paper \cite{Idziaszek2009} we have studied the
basic properties of atom-ion scattering and bound states, using
${}^{40}\mrm{Ca}^{+}\mrm{-}{}^{23}\mrm{Na}$ \cite{Makarov2003} as a
reference system. We have developed an effective atom-ion collision
model by applying the multichannel quantum defect theory (MQDT)
\cite{Seaton1983,Greene1982,Mies} to the polarization potential,
which scales as $r^{-4}$ at large atom-ion distance $r$. We have
verified that MQDT model predicts very accurately all atom-ion
scattering properties at ultracold temperatures by comparing to
numerical close-coupled calculations performed on {\it ab initio}
${}^{40}\mrm{Ca}^{+}\mrm{-}{}^{23}\mrm{Na}$ potentials. In the
literature, the MQDT approach has already been applied to describe scattering
and bound states in electron-ion core \cite{Seaton1983}, electron-atom
\cite{Watanabe1980} and neutral atom systems \cite{Gao2005}. Our model of
atom-ion collisions combines knowledge of the analytical solutions for the
$r^{-4}$ asymptotic potential \cite{Watanabe1980,Vogt1954,Spector1964}
with the idea of a frame transformation
\cite{Fano1970,Rau1971,Burke1998,Gao2005,Hanna2009,Hanna2010}. The
latter, applied at small distances allows a reduction of the number
of quantum-defect parameters needed to represent the effect of the
short-range interaction potential.

In \cite{Idziaszek2009} we have applied MQDT to the study of magnetic
Feshbach resonances and of radiative charge exchange processes,
including the effects of Feshbach and shape resonances. In this paper
we present a detailed derivation of our atom-ion model, based on the
Mies formulation of MQDT \cite{Mies}. We begin by introducing the
quantum-defect approach, and discuss the properties of analytical
solutions of the $r^{-4}$ asymptotic potential. We derive asymptotic
expansions of the model quantum-defect functions, valid for energies
below a characteristic quantity $E^\ast$ defined in terms of the atom-ion
interaction strength. For energies above $E^\ast$, an efficient numerical
algorithm to determine the characteristic exponent and the solutions
of the Mathieu equation needed for the MQDT is proposed. Knowledge of
the quantum defect parameters is crucial for understanding atom-ion
collisions in the ultracold domain, and allows us to derive several
non trivial analytical results concerning the behavior of weakly
bound states, $s$-wave magnetic Feshbach resonances, and radiative
charge-transfer probabilities. In comparison to the model presented in
Ref.~\cite{Idziaszek2009} we introduce an angular-momentum-dependent
correction that improves the accuracy of MQDT for high order partial
waves. This modification allows the regime of applicability of our
model to be extended up to mK temperatures. Using this revised model
and numerical close-coupled calculations, the effect of finite energy
Feshbach resonances on elastic and charge-transfer collision rates is
investigated as a function of the magnetic field.

In the present paper we also include details on the semiclassical
description of radiative charge-transfer rates. A combination of the
semiclassical approximation and MQDT scaling functions allows us to
represent the charge-exchange loss rate as the product between the
classical Langevin result and a universal quantum correction dependent
only on a single parameter, the short-range phase. In \cite{Idziaszek2009}
we discovered that the main product of a charge-transfer process for
${}^{40}\mrm{Ca}^{+}\mrm{-}{}^{23}\mrm{Na}$ are molecular ions. Here
we present a more detailed analysis by analyzing the population of
vibrational states formed after the electron transfer. A semiclassical
formula describing this distribution is derived. The very good
accuracy of our MQDT analytical results is tested by comparison with
the numerical solution of the close-coupled Schr\"odinger equation for
the ${}^{40}\mrm{Ca}^{+}\mrm{-}{}^{23}\mrm{Na}$ system.

The paper is organized as follows. Section~II introduces the MQDT
formalism. Section III discusses the properties of scattering and bound
states for a single channel potential, based on the analytical solutions
for the polarization potential. In particular, section~III.A derives
the analytical solutions, section~III.B presents the results of the
scattering states and analyzes the applicability of the semiclassical
description. Section~III.C presents the analytical results for weakly
bound states and section Section~III.D studies the low-energy behavior of
the MQDT functions. The idea of the frame transformation is discussed
in Section IV, using a particular example of an alkali atom and an
alkali-earth ion. Section~IV.A generalizes the frame transformation to
the presence of magnetic field, and Section~IV.B derives the corrections
to the quantum-defect matrix for higher partial waves. Section~V
describes the radiative charge transfer: the perturbative approach
based on the Fermi golden rule, and the semiclassical description of
the charge-transfer probability at short range. The theoretical model
is applied to the ${}^{40}\mrm{Ca}^{+}\mrm{-}{}^{23}\mrm{Na}$ system in
Section~VI. Section~VII presents the conclusions, and five appendices
give some technical details related to the derivation of the analytical
solutions and the multi-channel calculations of the radiative charge
transfer.

\section{Multichannel formalism and quantum defect theory}

In this section we summarize the basic formalism of the multichannel
quantum defect theory, adopting the formulation introduced by Mies
in Ref.~\cite{Mies}. Atom-ion collisions are described by a $N$-channel
close-coupled radial Schr\"odinger equation

\begin{equation}
\label{RadialSchr}
\frac{\partial^2 \mbf{F}}{\partial r^2} + \frac{2 \mu}{\hbar^2}\left(E - \mbf{W}(r) \right) \mbf{F}(r) = 0.
\end{equation}
Here $\mu=m_i m_a/(m_i+m_a)$ denotes the reduced mass, and $\mbf{W}(r)$ and
$\mbf{F}(r)$ are $N \times N$ matrices representing the interaction and the
radial solutions, respectively. A general solution to the $N$-channel scattering problem is given by a set of $N$ linearly independent wave functions
\begin{equation}
\label{Psi}
\Psi_i(\mbf{r}) = \sum_{j=1}^{N} |\Phi_j\rangle Y_{\ell_j}(\mbf{\hat{r}}) F_{ji}(r)/r, \quad i=1,\ldots,N
\end{equation}
where $|\Phi_j\rangle$ are channels states describing the internal spin degrees of freedom,
$Y_{\ell_j}(\hat{\mbf{r}})$ denotes the angular part of the solution (spherical harmonic) for the channel $j$.
The interaction matrix is asymptotically diagonal
\begin{equation}
\label{Wasympt}
W_{ij}(r) \stackrel{r \rightarrow \infty}{\longrightarrow} \left[ E_i^{\infty} +\frac{\hbar^2 \ell_i(\ell_i+1)}{2 \mu r^2} - \frac{C_{4}}{r^4}
\right]\delta_{ij}
\end{equation}
where $E_i^{\infty}$ are the threshold energies for the molecular
dissociation, $\ell_i$ is the partial wave quantum number of channel
$i$, $C_{4} = \alpha e^2/2$ with $\alpha$ denoting the static
dipolar polarizability of the atom and $e$ is the ion charge. Here we
neglect the contribution of the higher-order dispersion terms to the
long-range potential, which give relative small corrections in
atom-ion scattering \cite{Idziaszek2009}.

Given the total energy $E$, the channel states can be classified as open
for $E>E_i^{\infty}$ or closed for $E<E_i^{\infty}$. In the former case
the asymptotic wave number $k_i = \sqrt{2 \mu (E-E_i^{\infty})/\hbar^2}$
is real and positive, while in the latter is purely imaginary $k_i =
e^{i \pi/2} |k_i|$. The solution matrix $\mbf{F}(r)$ can be split into
blocks
\begin{equation}
\mbf{F}(r) = \left( \begin{array}{cc}
\mbf{F}_\mrm{oo}(r) & \mbf{F}_\mrm{oc}(r) \\
\mbf{F}_\mrm{co}(r) & \mbf{F}_\mrm{cc}(r)
\end{array} \right)
\end{equation}
corresponding to $N_o$ open and $N_c$ closed channels. Imposing
appropriate asymptotic boundary conditions on the closed channel
components, $F_{ij} \rightarrow 0$ ($r \rightarrow \infty$) for $i=
N_o+1,\ldots,N$, the physically meaningful part of $\mbf{F}(r)$ is
contained in the block $(N \times N_o)$. The observable properties of
the atom-ion system result from the asymptotic behavior of the open-open
block $\mbf{F}_\mrm{oo}(r)$, which at large distances yields the reactance
matrix $\mbf{K}_\mrm{oo}$
\begin{equation}
\label{Foo}
\mbf{F}_\mrm{oo}(r) \stackrel{r \rightarrow \infty}{\longrightarrow}
\left[\mbf{J}(r) - \mbf{N}(r) \mbf{K}_\mrm{oo} \right] \mbf{A}_\mrm{oo}.
\end{equation}
Here, $J_{ij}(r) \rightarrow \delta_{ij} \sin(k_i r - \ell_i
\pi/2)/\sqrt{k_i}$ and $N_{ij}(r) \rightarrow - \delta_{ij} \cos(k_i r -
\ell_i \pi/2)/\sqrt{k_i}$ ($r \rightarrow \infty$) exhibit asymptotic
behavior associated with the spherical Bessel functions $j_\ell(kr)$
and $n_\ell(kr)$. The constant matrix $\mbf{A}_\mrm{oo}$ depends on the
boundary conditions at $r \rightarrow \infty$. In particular, the choice
$\mbf{A}_\mrm{oo} = (\mbf{1} - i \mbf{K}_\mrm{oo})^{-1}$ corresponds to
the usual incoming-wave boundary conditions, where the amplitude of the
outgoing wave is determined by the scattering matrix $\mbf{S}_\mrm{oo} =
(\mbf{1} + i \mbf{K}_\mrm{oo})(\mbf{1} - i \mbf{K}_\mrm{oo})^{-1}$.

The basic idea of the quantum-defect theory is to introduce a set of
parameters describing the short-range behavior of the wave function,
that weakly depend on total energy $E$, and can be used to predict
the system properties as $E$  crosses the dissociation thresholds of
the individual channels. As we will show later, in the ultracold domain
the quantum defect parameters for atom-ion collisions are also weakly
dependent on the relative orbital angular momentum $\ell$ and can be
taken as constant, at least for the lowest partial waves.

We now specialize the formulation of MQDT developed by Mies for atomic
collisions \cite{Mies} to our atom-ion system. The starting point is
the choice of a set of reference potentials $\{V_j(r)\}$ that should
reproduce the asymptotic behavior of the interaction matrix at large
distances $V_j(r) \stackrel{r \rightarrow \infty}{\longrightarrow}
W_{jj}(r)$ but can otherwise be arbitrary. One associates to the
reference potentials $V_i(r)$ a pair of linearly independent solutions
$\hat{f}_i(r)$ and $\hat{g}_i(r)$ that have short-range WKB-like
normalization

\begin{subequations}
\label{fghat}
\begin{align}
\hat{f}_{i}(r) & = \alpha_i(r) \sin \beta_i (r),\\
\hat{g}_{i}(r) & = \alpha_i(r) \cos \beta_i (r).
\end{align}
\end{subequations}

The amplitude $\alpha_i(r)$ fulfills the inhomogeneous Milne equation:
$[d^2/dr^2 + k_i(r)^2] \alpha_i(r) = \alpha_i^{-3}(r)$ \cite{MilnePR1930},
with the local wavevector $k_i(r) = \sqrt{2 \mu (E - V_i(R)}/\hbar$,
while the phase $d\beta_i/dr = 1/\alpha_i^2$.
Since the reference potentials $\{V_j(r)\}$ reproduce the asymptotic
behavior of $\mbf{W}(r)$, the exact solution to Eq.\eqref{RadialSchr}
can be expressed at large distances in terms of pair of functions
$\hat{\mbf{f}}(r) \equiv \{\delta_{ij} \hat{f}_{i}(r)\}$ and
$\hat{\mbf{g}}(r) \equiv \{\delta_{ij} \hat{g}_{i}(r)\}$ \begin{equation}
\mbf{F}(r) \stackrel{r \rightarrow \infty}{\longrightarrow}
\left[\hat{\mbf{f}}(r) + \hat{\mbf{g}}(r) \mbf{Y} \right] \hat{\mbf{A}}
\end{equation} Here, $\mbf{Y}$ is the quantum-defect matrix that play
a central role in the MQDT analysis. In contrast to the $\mbf{S}$ and
$\mbf{K}$ scattering matrices, $\mbf{Y}$ remains analytic across the
thresholds, and has in general only a weak dependence on energy.

The observable properties depend on the asymptotic behavior of the
solutions at large distance. It is therefore convenient to introduce
another pair of solutions, $f_i(r)$ and $g_i(r)$, together with the
physically well-behaved solution $\phi_i(r)$ for the closed channels,
that have energy-like normalization as $r \rightarrow \infty$
\begin{subequations}
\label{fg}
\begin{align}
f_i(r) & \cong k_i^{-1/2}\sin (k_ir-\ell_i\pi/2+\xi_i),& E \geq E_i^\infty \\
g_i(r) & \cong k_i^{-1/2}\cos (k_ir-\ell_i\pi/2+\xi_i),& E \geq E_i^\infty\\
\phi_i(r) & \cong \frac{1}{2} k_i^{-1/2} e^{- |k_i| r},& E \leq E_i^\infty .
\end{align}
\end{subequations}

The factor $\frac12$ in the last equation is introduced for convenience
in order to simplify the value of Wronskian \cite{Mies}. Here, $\xi_i$
is the scattering phase shift for a channel $i$. The MQDT functions
$C_i(E)$, $\tan \lambda_i (E)$ and $\nu_i(E)$ relate the solutions
\eqref{fg} to \eqref{fghat}, and are defined as follows
\begin{subequations}
\label{MQDTfunctions}
\begin{align}
\label{f}
f_i(r) & = C^{-1}_i(E) \hat{f}_i(r), & E \geq E_i^\infty \\
\label{g}
g_i(r) & = C_i(E) \left[ \hat{g}_i(r) + \tan \lambda_i (E) \hat{f}_i(r) \right], & E \geq E_i^\infty\\
\label{phi}
\phi_i(r) & = {\cal N}_i(E) \Big[ \cos \nu_i(E) \hat{f}_i(r) - \sin \nu_i(E) \hat{g}_i(r) \Big],& E \leq E_i^\infty
\end{align}
\end{subequations}

Now, when all the channels are open: $\forall_i E \geq E_i^\infty$,
the solution $\mbf{F}(r)$ at $r \rightarrow \infty$ can be expressed in
terms of $\mbf{f}(r) \equiv \{\delta_{ij} f_{i}(r)\}$ and $\mbf{g}(r)
\equiv \{\delta_{ij} g_{i}(r)\}$ \begin{equation} \mbf{F}(r) \stackrel{r
\rightarrow \infty}{\longrightarrow} \left[\mbf{f}(r) + \mbf{g}(r)
\mbf{R}(E)\right] \mbf{A} \end{equation} Using the relationships
\eqref{f}-\eqref{phi} one can show that \begin{equation}\label{R}
\mbf{R}(E) = \mbf{C}^{-1}(E) \left[\mbf{Y}^{-1}(E) - \tan {\bm \lambda}(E)
\right]^{-1} \mbf{C}(E) \end{equation} where $\mbf{C}(E) \equiv
\{\delta_{ij} C_{i}(E)\}$ and ${\bm \lambda}(E) \equiv \{\delta_{ij}
\lambda_{i}(E)\}$.
At high energies the WKB approximation is valid at all distances, and
the functions $\hat{f}$, $\hat{g}$ become identical to $f$ and $g$,
respectively. Therefore, in this situation the MQDT parameters behave
like $C_i(E) \rightarrow 1$ and  $\tan \lambda_i(E) \rightarrow 0$,
and Eq.~\eqref{R} reduces to $\mbf{R}(E) \cong \mbf{Y}(E)$.

Next, with the help of Eq.~\eqref{Foo} one can relate the reactance matrix
$\mbf{K}$ and the scattering matrix $\mbf{S}$ to the matrix $\mbf{R}$
\begin{align}
\label{K}
\mbf{K} & = \left[\sin({\bm \xi}) + \cos({\bm \xi}) \mbf{R} \right]
\left[\cos({\bm \xi}) - \sin({\bm \xi})\mbf{R} \right]^{-1}, \\
\label{S}
\mbf{S} & = e^{i {\bm \xi}} \left[ \mbf{1} + i \mbf{R} \right]
\left[\mbf{1} - i \mbf{R} \right]^{-1} e^{i {\bm \xi}},
\end{align}
with ${\bm \xi}(E) \equiv \{\delta_{ij} \xi_{i}(E)\}$.

The current derivation has to be modified in the presence of
closed channels: $E < E_i^\infty$ for $i=N_o+1,\ldots,N$. In this
case, we impose the  requirement that the wave function of the closed
channels decays exponentially for large $r$, i.e. the closed channel
wave functions are proportional to $\phi_i(r)$. This results in the
renormalization of the open-open block of the quantum-defect matrix
\cite{Mies} \begin{equation} \mbf{\bar{Y}}_\mrm{oo} = \mbf{Y}_\mrm{oo}
- \mbf{Y}_\mrm{oc} \left[ \tan({\bm \nu}_\mrm{cc}) + \mbf{Y}_\mrm{cc}
\right]^{-1} \mbf{Y}_\mrm{co}, \end{equation} where ${\bm \nu}(E)
\equiv \{\delta_{ij} \nu_{i}(E)\}$ The scattering matrices can now be
calculated from Eqs.~\eqref{R}-\eqref{S}, applied only to the open-open
block, where one substitutes $\mbf{\bar{Y}}_\mrm{oo}$ in place of
$\mbf{Y}_\mrm{oo}$. Finally, when all the channels are closed ($N_c =
N$), the wave functions of all the channels must be proportional to
$\phi_i(r)$ at large $r$, and the energies of the bound states are
determined by the condition
\begin{equation}
\label{MultChEb}
\left|\mbf{Y}(E) + \tan {\bm \nu}(E)\right| = 0.
\end{equation}

\section{Long-range atom-ion interaction}

\subsection{Analytical solutions}

Here we focus on the Schr\"odinger equation for a single channel where
we include only the long-range part of the atom-ion interaction and the
centrifugal barrier for partial wave $\ell$
\begin{equation}
\label{RadSchr}
\frac{\partial^2 F}{\partial r^2} + \frac{2 \mu}{\hbar^2}
\left(E-\frac{\hbar^2 \ell(\ell+1)}{2 \mu r^2} - \frac{C_4}{r^4} \right)F(r) = 0.
\end{equation}

\begin{figure}
\begin{indented}
\item[]\includegraphics[width=8cm,clip]{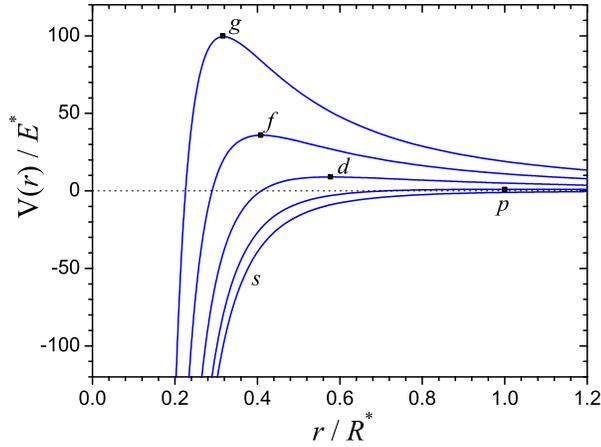}
\end{indented}
\caption{The long-range atom-ion potentials for the few lowest partial waves scaled by the characteristic distance $R^\ast$ and the characteristic energy $E^\ast$.
	 \label{fig:V}}	
\end{figure}

In the following we will work in dimensionless units, where the
length is expressed in units of $R^\ast \equiv \sqrt{ 2 C_4 \mu
/\hbar^2}$ and energy in units of $E^{\ast} \equiv \hbar^2/\left[2 \mu
(R^\ast)^2\right]$. Table~\ref{Tab} presents the characteristic lengths
$R^\ast$ and energies $E^\ast$ for some sample combination of alkali
atoms and alkali-earth ions. Fig.~\ref{fig:V} shows the long-range
atom-ion potentials for the lowest partial waves, where the squares mark the
top of centrifugal barriers occurring at $r_\mrm{max} =
\sqrt{2}/\sqrt{\ell(\ell+1)} R^{\ast}$ with energy $E_\mrm{max} =
\frac14 \ell^2 (\ell+1)^2 E^{\ast}$. The Eq.\eqref{RadSchr} in characteristic
units of $R^\ast$ and $E^{\ast}$ takes the form
\begin{equation}
\label{RadSchr1}
\frac{\partial^2 F}{\partial r^2} + \left( E-\frac{\ell(\ell+1)}{r^2} + \frac{1}{r^4} \right)F(r) = 0.
\end{equation}
\Table{\label{Tab} Characteristic distance $R^{\ast}$ and characteristic energy $E^{\ast}$ for some selected atom-ion systems.}
\br
 & $R^{\ast}$(units of $a_0$) & $E^{\ast}/h$ (kHz) \\
\mr
$^{40}$Ca$^{+}$ + $^{23}$Na & 2081 & 28.56   \\
$^{40}$Ca$^{+}$ + $^{87}$Rb & 3989 & \04.143 \\
$^{135}$Ba$^{+}$ + $^{87}$Rb & 5544 & \01.111   \\
$^{172}$Yb$^{+}$ + $^{87}$Rb & 5793 & \00.9313  \\
\br
\endTable
This equation can be solved analytically \cite{Vogt1954,OMalleyJMP1961,Spector1964} by substituting $F(r) = \psi(r) r ^{1/2}$ and $r= e^z E^{-1/4}$, which yields the Mathieu's equation of imaginary argument
\begin{equation}
\label{Mathieu}
\frac{d^2 \psi}{d z^2} - \left[ a - 2 q \cosh 2 z \right] \psi = 0.
\end{equation}
with $a=(\ell+{\tst \frac 12})^2$ and $q=\sqrt{E}$. Although
the Mathieu's equation is well known in mathematical physics (see
e.g. \cite{Abramowitz,Erdelyi}), we include a brief derivation of its
basic properties in the Appendix~\ref{App:Mathieu}. The Appendix also discusses
the basic methods we applied for its numerical solution.

We denote two linearly independent solutions to Eq.~\eqref{RadSchr1} by
$T_\nu(r)$ and $T_{-\nu}(r)$, where $\nu$ is the complex number called
the characteristic exponent (see Appendix~\ref{App:Mathieu}
for definition).  In view of the asymptotic properties of $T_\nu(z)$
discussed in  Appendix~\ref{App:Mathieu}), we can construct two solutions
$\hat{f}(r)$ and $\hat{g}(r)$ defined by Eqs.~\eqref{fghat}
\begin{align}
\hat{f}(r) = & A_{\nu}(\phi) T_\nu(r) + A_{-\nu}(\phi) T_{-\nu}(r) \\
\hat{g}(r) = & A_{\nu}(\phi+\pi/2) T_\nu(r) + A_{-\nu}(\phi+\pi/2) T_{-\nu}(r),
\end{align}
where
\begin{equation}
A_{\nu}(\phi)= \frac{\sin(\phi - \nu \pi/2 + \pi/4)}{\sin(\pi \nu)}
\end{equation}
and $\phi$ is some parameter that can be interpreted as the short-range
phase. As can be easily verified, $\hat{f}(r)$ and $\hat{g}(r)$ exhibit
at distances $r \ll 1$ semiclassical behavior given by Eqs.\eqref{fghat}
with $\alpha(r) \cong r$ and $\beta(r) \cong -1/r$
\begin{subequations}
\label{fghatS}
\begin{align}
\label{fhatS}
\hat{f}(r) &\stackrel{r \rightarrow 0}{\longrightarrow} r \sin(-1/r + \phi), \\
\label{ghatS}
\hat{g}(r) &\stackrel{r \rightarrow 0}{\longrightarrow} r \cos(-1/r + \phi),
\end{align}
\end{subequations}

The asymptotic behavior given by Eqs.\eqref{fghatS} can be also obtained
by solving directly Eq.~\eqref{RadSchr1} with centrifugal and energy
terms neglected. Since $r \sin(-1/r + \phi)$ is an exact solution of
\eqref{RadSchr} for $\ell=0$ and $E=0$, it can be used to derive a
simple relation between the short-range phase $\phi$ and the $s$-wave
scattering length
\begin{equation}
\label{asc}
a = R^{\ast} \cot \phi,
\end{equation}
that follows from the asymptotic behavior: $F(r) \rightarrow r-a$ ($r \rightarrow \infty$) at $E=0$.

At large distances, the solutions $\hat{f}(r)$ and $\hat{g}(r)$
behaves according to
\begin{subequations}
\label{fghatL}
\begin{align}
\label{fhatL}
\hat{f}(r) \stackrel{r \rightarrow \infty}{\longrightarrow} &
C_{\nu}(\phi)\sqrt{q} r j_\ell(q r) -
D_{\nu}(\phi)\sqrt{q} r n_\ell(q r), \\
\label{ghatL}
\hat{g}(r) \stackrel{r \rightarrow \infty}{\longrightarrow} &
C_{\nu}\left(\phi+\frac\pi2\right)\sqrt{q} r j_\ell(q r) - D_{\nu}\left(\phi+\frac\pi2\right)\sqrt{q} r n_\ell(q r),
\end{align}
\end{subequations}
where
\begin{subequations}
\begin{align}
C_{\nu}(\phi) = & A_{\nu}(\phi) m_{\nu} \cos\eta - (-1)^\ell A_{-\nu}(\phi) m_{-\nu} \sin\eta,\\
D_{\nu}(\phi) = & (-1)^\ell
A_{-\nu}(\phi) m_{-\nu} \cos\eta - A_{\nu}(\phi) m_{\nu} \sin\eta.
\end{align}
\end{subequations}
Here, $\eta = \frac \pi 2 (\nu -\ell -\frac 1 2)$, $m_{\nu} = S_{\nu}
(4/q)^\nu$, and $S_\nu$ is a function of  $\nu$, that is defined in
terms of a continued fraction (see Appendix~\ref{App:Mathieu} for the
definition).

Making use of the asymptotic behavior \eqref{fghatL} one can find the phase shift $\xi$
\begin{equation}
\label{tanxi}
\tan \xi = D_{\nu}(\phi)/C_{\nu}(\phi),
\end{equation}
and all MQDT functions defined in \eqref{f}-\eqref{phi}
\begin{flalign}
\label{C}
C(E) & = C_{\nu}(\phi) /\cos \xi\\
\label{tanl}
\tan \lambda(E) & = C^{-2}(E) \tan^{-1}(\xi-\tilde{\xi}) \\
\label{tanniu}
\tan \nu(E) & = \frac{A_{\nu}(\phi)+A_{-\nu}(\phi) S_{\nu}^{-2}
\left(\chi/4\right)^{2\nu}}{A_{\nu}(\phi+\frac\pi2)+A_{-\nu}(\phi+\frac\pi2) S_{\nu}^{-2} (\chi/4)^{2\nu}}
\end{flalign}
where $\chi=\sqrt{-E}$, and $\tilde{\xi}$ is the phase shift of a second short-range normalized solution $\hat{g}(r)$
\begin{equation}
\tan \tilde{\xi} = \frac{D_{\nu}(\phi+\pi/2)}{C_{\nu}(\phi+\pi/2)},
\end{equation}

Finally we point out that formulas \eqref{fghatL} describing the
asymptotic behavior of $\hat{f}(r)$ and $\hat{g}(r)$ can be used to
determine the scattering matrix $\mbf{K}$ directly from the quantum-defect
matrix $\mbf{Y}$
\begin{equation}
\mbf{K} = \left[\mbf{Y}(\phi)\mbf{C}_\nu(\phi+\pi/2) - \mbf{D}_\nu(\phi+ \pi/2)\right]^{-1}
\left[ \mbf{Y}(\phi) \mbf{C}_\nu(\phi) - \mbf{D}_\nu(\phi)\right],
\end{equation}
where ${\bm C}_\nu(\phi) \equiv \{\delta_{ij} C_{\nu(i)}(\phi)\}$ and
${\bm D}_\nu(\phi) \equiv \{\delta_{ij} D_{\nu(i)}(\phi)\}$. We note that
in principle the parametrization in terms of the quantum-defect matrix
$\mbf{Y}(\phi)$ depends on the short-range phase $\phi$ that is fixed by
the choice of the reference potentials. In contrast, the scattering matrix
$\mbf{K}$ will depend only on the actual scattering lengths associated
with the scattering channels and should be independent of a particular
set of reference functions that determine specific $\phi$ for that choice.

\subsection{Scattering states and semiclassical approximation}
\label{Sec:scatt}

Using Eq.\eqref{tanxi} and the small-$q$ expansions presented
in Appendix~\ref{App:ExpSmallQ} one can derive the well known
threshold behavior of the phase shifts in the polarization potential
\cite{OMalleyJMP1961,Spector1964}
\begin{flalign}
\label{xil}
\tan \xi_0(q) & =  -a q - {\tst \frac{\pi}{3}} q^2 + {\cal O}(q^3), \\
\label{xil1}
\tan \xi_{\ell}(q) & = \frac{\pi q^2}{8(\ell-\frac12)(\ell+\frac12)(\ell+\frac32)} + {\cal O}(q^3), \qquad \ell >0,
\end{flalign}

\begin{figure}
\begin{indented}
\item[]\includegraphics[width=8cm,clip]{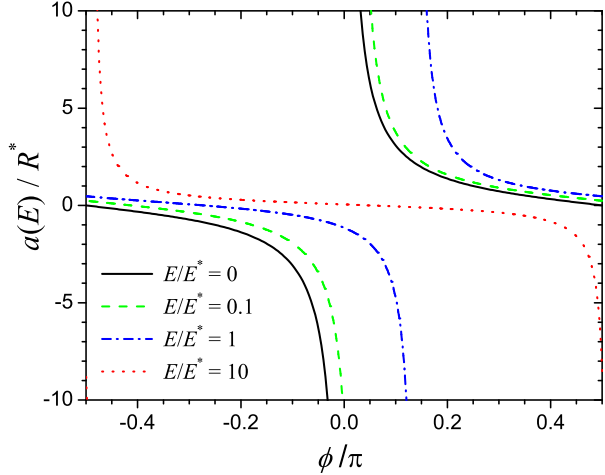}
\end{indented}
\caption{
\label{fig:asc}
Energy-dependent scattering length $a(k)$ as a function of
the short-range phase $\phi$ for different values of the collision energy.}
\end{figure}

Here, we have used the relation \eqref{asc} between the short-range phase
$\varphi$ and the scattering length $a$ of the reference potentials,
in order to express all the formulas in terms of $a$.  Because of the
long-range character of the polarization potential, the scattering length
can be defined only for $\ell=0$, while for $\ell>0$ the threshold
behavior of the phase shifts is dominated by the contribution of the
polarization potential. Fig.~\ref{fig:asc} shows the energy-dependent
scattering length $a(k) = - \tan \xi_0 / k$ as a function of $\phi$
for diferent values of the energy. The case $k=0$ corresponds to $a =
R^{\ast} \cot \phi$.

\begin{figure}
\begin{indented}
\item[]\includegraphics[width=8cm,clip]{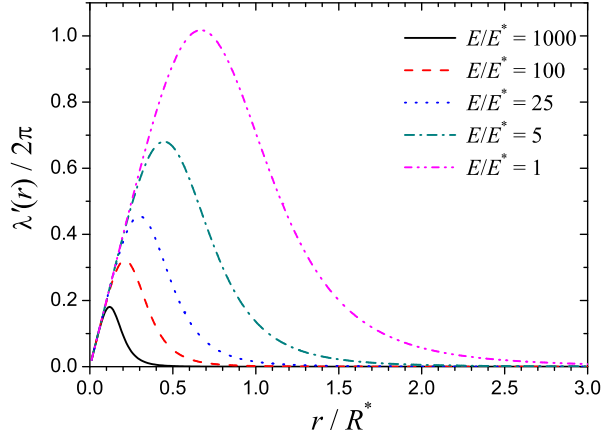}
\end{indented}
\caption{
\label{fig:WKBcond}
Derivative of the local de Broglie wavelength $\lambda(r)$ giving the condition for the applicability of the WKB approximation plotted for different energies.}
\end{figure}

\begin{figure}
\begin{indented}
\item[]\includegraphics[width=8cm,clip]{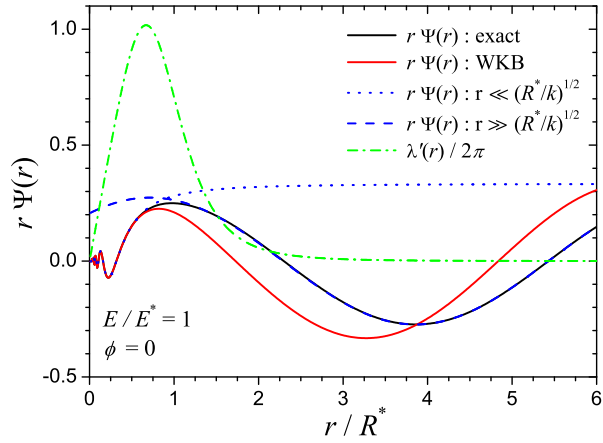}
\end{indented}
\caption{
\label{fig:wf1}
Exact wave function (black solid line) compared with the WKB wave function
(red solid line) and with small and large $r$ asymptotic approximations,
given by Eqs.~\eqref{fhatS} and \eqref{fhatL}, respectively (blue
dotted and dashed lines), for energy $k R^{\ast} = 1$ and $\phi =
0$. The figure also shows the derivative of the local wavelength
$\lambda(r)$ (green dot-dashed line), representing the condition for the applicability of the
WKB approximation.}
\end{figure}

\begin{figure}
\begin{indented}
\item[]\includegraphics[width=8cm,clip]{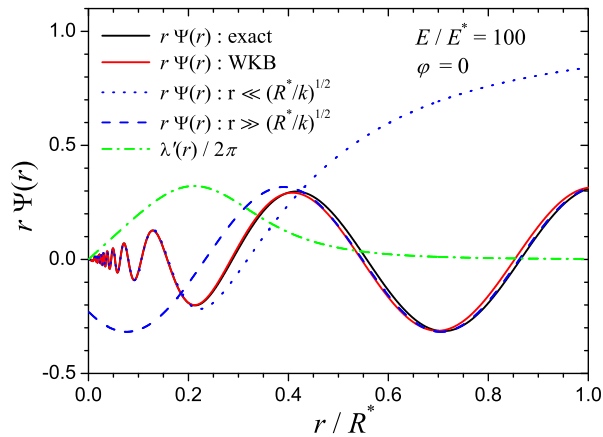}
\end{indented}
\caption{
\label{fig:wf100}
The same as Fig.~\ref{fig:wf1} but for energy $E = 100 E^{\ast}$.}
\end{figure}

At sufficiently high energies the scattering can be described using
the semiclassical approximation. To identify the crossover from the
quantum to the semiclassical regime, we plot in Fig.~\ref{fig:WKBcond}
the quantity $\lambda^\prime(r)/{2 \pi}$, where $\lambda(r)$ is the
local de Broglie wavelength. A necessary condition for the applicability
of the WKB approximation, expressed in terms of the local wavelength,
is $\lambda^\prime(r) \ll 2 \pi$. We observe that this condition
is first violated at distances comparable to $R^{\ast}$, while the WKB
approximation remains valid at small and large distances. We have
verified that to a good approximation the wave function can be calculated
within the semiclassical approximation when $\lambda^\prime(r)/{2 \pi}
\lesssim 1/2$, a condition fulfilled for energies $E \gtrsim 25 E^{\ast}$.
Figs.~\ref{fig:wf1} and \ref{fig:wf100} compare predictions of the
WKB method with the exact wave functions $\hat{f}(r)$ for two values
of energy: $E = E^{\ast}$ and  $E = 100 E^{\ast}$. In addition they
show the small-$r$ and large-$r$ asymptotic formulas \eqref{fhatS} and
\eqref{fhatL}, respectively, and derivative of the local wavelength:
$\lambda^\prime(r)/{2 \pi}$.

\subsection{Bound states}

\begin{figure}
\begin{indented}
\item[]\includegraphics[width=8cm,clip]{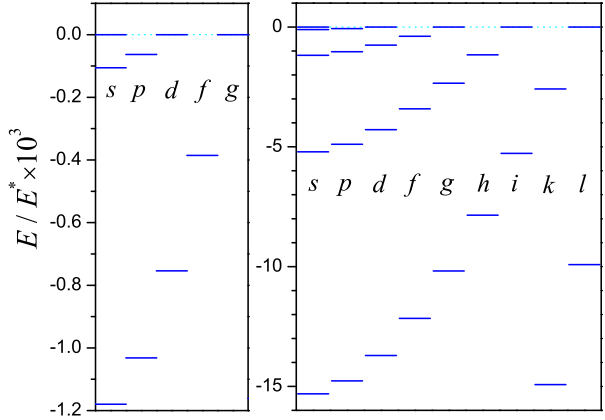}
\end{indented}
\caption{
\label{fig:ainf}
Energies of bound states for $a = \pm \infty$ and for different partial waves.}
\end{figure}

For negative energies the exponentially decaying solution is given by a
linear combination of $\hat{f}(r)$ and $\hat{g}(r)$ [see Eq.~\eqref{phi}]
with $\nu$ given by \eqref{tanniu}. In the quantum-defect approach the bound
state spectrum is determined by the condition $Y +\tan \nu(E)=0$, which
is a single-channel version of Eq.~\eqref{MultChEb}. In the presence of a
single channel one can take $Y=0$, which is satisfied when the reference
potential is chosen to reproduce the same scattering length as the real
physical potential. This yields the condition $\tan \nu(E)=0$. In this
case the bound state wave function $\phi(r)$ of Eq.~\eqref{phi} contains
only the $\hat{f}(r)$ component, with the short-range phase $\phi$ given in
terms of the scattering length by Eq.~\eqref{asc}. The condition determining the
energy of a bound state can be rewritten as
\begin{equation}
\label{bound}
\frac{\sin\left[\frac{\pi}{2} \nu(E) - \phi - \frac{\pi}{4} \right]}
{\cos\left[\frac{\pi}{2} \nu(E) + \phi - \frac{\pi}{4} \right]}
= - \left(-\frac{E}{16}\right)^{\nu(E)} S_\nu^{-2}(E),
\end{equation}
with $S_\nu(E)$ defined in the Appendix~\ref{App:TwoSol}.
Fig.\ref{fig:ebound} shows energies of the bound states versus the
short-range phase $\phi$ for the lowest partial waves. We observe that
for $\ell$ even the bound states disappear at the threshold at $\phi=0$
($a=\pm \infty$), while for odd $\ell$ this happens for $\phi=\pi/2$
$(a=0$). Thus, for a polarization potential there is an $\ell=2$
periodicity in the values of $\phi$ at the threshold, which is reminiscent
of $\ell=4$ periodicity for van der Waals potential \cite{GaoPRA2000}. The
energies of the bound states for $a=\pm \infty$ and for different partial
waves are shown in Fig.~\eqref{fig:ainf}. For even $\ell$ this defines
characteristic energy bins, which determine the positions of the last
bound states in the spectrum. For instance, the last $s$-wave bound
state is located in the energy range $E/E^\ast=[-106,0]$

\begin{figure}
\begin{indented}
\item[]\includegraphics[width=8cm,clip]{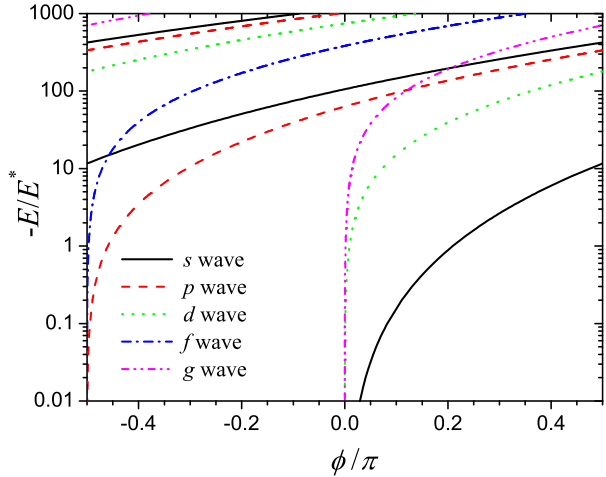}
\end{indented}
\caption{
\label{fig:ebound}
Energies of bound states as a function of the short-range phase $\phi$ for few lowest partial waves.}
\end{figure}

\begin{figure}
\begin{indented}
\item[]\includegraphics[width=8cm,clip]{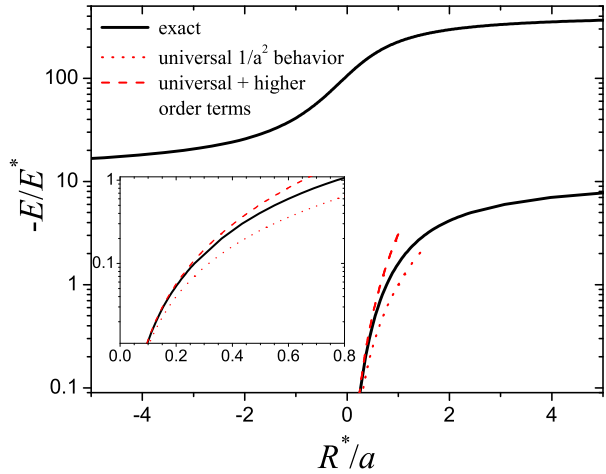}
\end{indented}
\caption{
\label{fig:ebounds}
Energies of $s$-wave bound states versus the inverse scattering length
$R^\ast/a$. Exact results of Eq.~\eqref{bound} (black solid line) are
compared with the universal law $E/E^{\ast} = - (R^{\ast}/a)^2$ (red dotted
line), and with the approximation \eqref{eblow} including terms up to
the order of $1/a^3$ (red dashed line).}
\end{figure}

Applying small $q$ expansions of $\nu$ and $S_\nu$ (see Appendix A)
one can expand the right-hand-side of Eq.~\eqref{bound}, which yields
for $s$-waves
\begin{equation}
\label{1aExpS}
\frac{1}{a} = \kappa - \frac{\pi}{3} \kappa^2 + {\cal O}(\kappa^3), \qquad \ell = 0,
\end{equation}
and for $p$-waves
\begin{equation}
\label{1aExpP}
a = - \frac{\pi}{15} \kappa^2 -\frac{\kappa^3}{9} + {\cal O}(\kappa^4), \qquad \ell = 1,
\end{equation}
with $\kappa = \sqrt{-E}$. Inverting the former equation, we can write the energy of $s$-wave bound states in powers of $1/a$
\begin{equation}
\label{eblow}
E = - \frac{1}{a^2} + \frac{2 \pi}{3} \frac{1}{a^3} + {\cal O}(1/a^4), \qquad \ell = 0.
\end{equation}
The first term on the right-hand-side is the universal energy of a
weakly bound state, the higher order term represents the correction
that is specific for $1/r^4$ potential. Fig.~\ref{fig:ebounds} shows
the binding energy for an $s$-wave bound state versus the inverse of
the scattering length. We observe that the range of the applicability of
the universal formula $E = -1/a^{2}$ is very narrow.  Inclusion of the
higher order correction in $1/a$ given by Eq.~\eqref{eblow} improves the
agreement with the exact one. In fact the approximation \eqref{eblow}
works reasonably for $|E| \lesssim 0.5 E^\ast$.

\subsection{Behavior of MQDT functions for polarization potential}
\label{Sec:MQDTfun}

\begin{figure}
\begin{indented}
\item[]\includegraphics[width=8.6cm,clip]{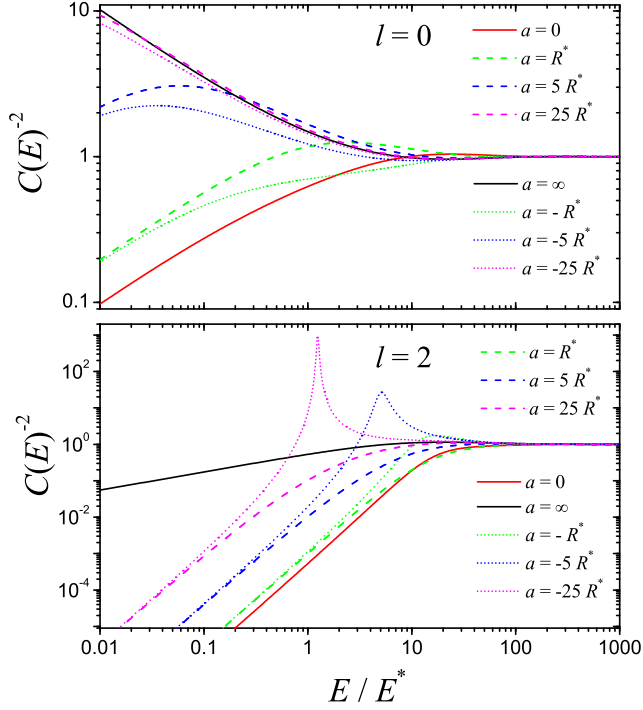}
\end{indented}
\caption{
\label{fig:CE}
$C(E)$ function versus the energy for partial waves $\ell=0$ (upper panel) and $\ell=2$ (bottom panel), calculated for different values of the scattering length $a$.}
\end{figure}

\begin{figure}
\begin{indented}
\item[]\includegraphics[width=8.6cm,clip]{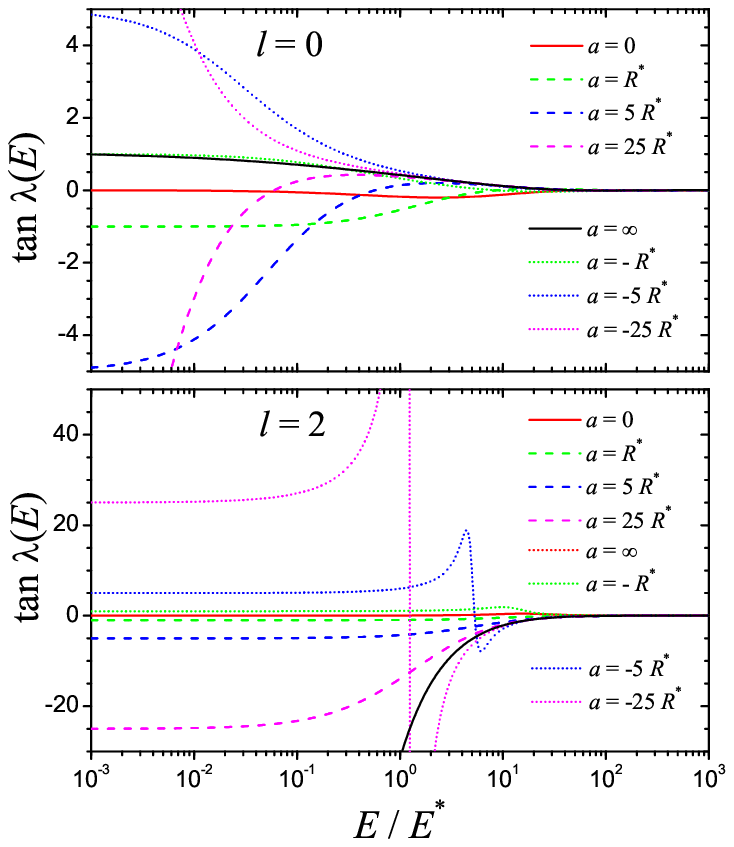}
\end{indented}
\caption{
\label{fig:TanL}
$\tan \lambda (E)$ function versus the energy for partial waves $\ell=0$ (upper panel) and $\ell=2$ (bottom panel), calculated for different values of the scattering length $a$.}
\end{figure}

\begin{figure}
\begin{indented}
\item[]\includegraphics[width=8.6cm,clip]{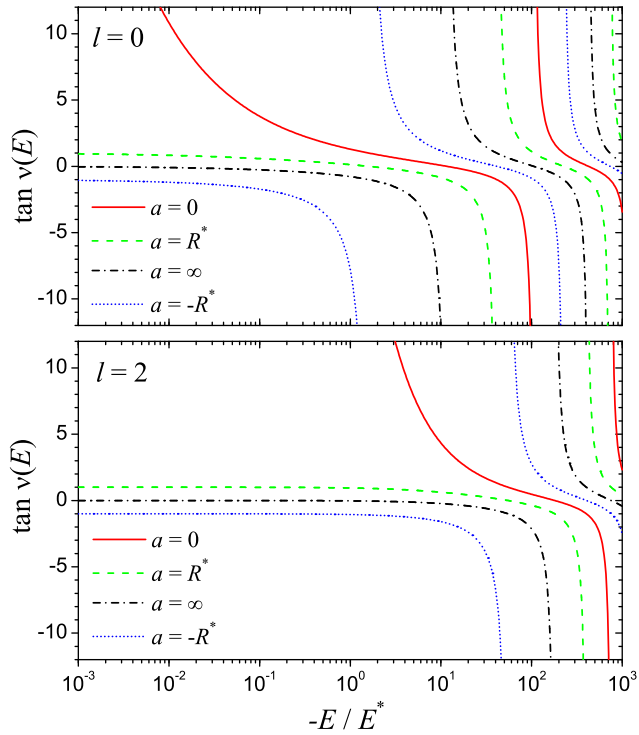}
\end{indented}
\caption{
\label{fig:TanNu}
$\tan \nu (E)$ function versus the energy for partial waves $\ell=0$ (upper panel) and $\ell=0$ (bottom panel), calculated for different values of the scattering length $a$.}
\end{figure}

Figs.~\ref{fig:CE}-\ref{fig:TanNu} show the $C(E)$, $\tan \lambda
(E)$ and $\tan \nu(E)$ functions for angular momenta $\ell=0$
and $2$ determined from Eqs.~\eqref{C}-\eqref{tanniu} for
different values of the $s$-wave scattering length $a$ i.e., choice
of $\phi$ in Fig.~\ref{fig:asc}. The calculations are performed under
the assumption that the short range phase $\phi$ does not depend on
$\ell$, such that all the MQDT functions can be parameterized with a single
$\ell$-independent $\phi$. Substituting the small-$q$ expansions given
in Appendix~\eqref{App:ExpSmallQ} into Eqs.~\eqref{C} and \eqref{tanl}
one obtains
\begin{subequations}
\label{smallq}
\begin{align}
\label{CEa}
C^{2}(E) & \stackrel{E \rightarrow 0^+}{\sim} \frac{\Gamma\left(\ell+\frac32\right)^2}{\Gamma\left(\frac12-\ell\right)^2} \sin^2\left(\phi+ \ell \frac{\pi}{2}\right) \left(\frac 4 q\right)^{2\ell+1}
 + \delta_{\ell,0} q \cos^2 \phi  + {\cal O}(q^2), \\
\label{TanLa}
\tan \lambda(E) & \stackrel{E \rightarrow 0^+}{\sim}
- \cot \left(\phi + \ell \frac{\pi}{2} \right)
+ \delta_{\ell,0} q^2 \frac{\cot \phi}{\sin^2 \phi}  + {\cal O}(q^3),\\
\tan \nu(E) & \stackrel{E \rightarrow 0^-}{\sim}
\tan \left(\phi + \ell \frac{\pi}{2} \right) + \delta_{\ell,0} \frac{\kappa(\kappa \tan \phi -1)}{\cos^2 \phi}  - \frac{\pi \kappa^2}{8(\ell-\frac12)(\ell+\frac12)(\ell+\frac32)} + {\cal O}(q^3).
\end{align}
\end{subequations}
Here, $\delta_{\ell,0}=1$ for $\ell=0$ and is zero otherwise. At
sufficiently large energy, the semiclassical solution begins to be
valid at all distances, implying $C(E) \rightarrow 1$ and  $\tan \lambda(E)
\rightarrow 0$. One can verify that the MQDT functions presented in
Figs.~\ref{fig:CE}-\ref{fig:TanNu} exhibit the small-$q$ and large-$q$
asymptotic behavior given by Eqs.~\eqref{smallq}. For nonzero angular
momenta one can observe the existence of sharp peaks in the function
$C(E)$, corresponding to shape resonances, that appear due to the presence
of quasi-bound states behind the centrifugal barrier.


\section{Frame transformation}
\label{Sec:Frame}

In the ultracold regime variations of the total energy $E$ are typically
much smaller than the depth of the potential at short range, where the matrix
$\mbf{Y}$ is defined. Therefore, it is justified to neglect the dependence
of $\mbf{Y}(E)$ on energy and to set $\mbf{Y}(E) \cong \mbf{Y}$. In this
way, the matrix $\mbf{Y}$ can be determined at a single value of energy, for
instance from the reaction or scattering matrices $\mbf{K}$ or $\mbf{S}$ by applying
formulas \eqref{K} and \eqref{S}, respectively. An alternative way is
to apply the frame transformation technique \cite{Gao1996,Burke1998}
that provides a very convenient way of parameterizing the short range
matrix $\mbf{Y}$ in terms of few parameters, e.g. scattering lengths.

In this paper we focus on the collisions of an alkali atom with an
alkali-earth ion in their electronic ground states. Hence, the asymptotic
channel states can be characterized by the hyperfine quantum numbers
$f_1$,$m_{f_1}$ and $f_2$,$m_{f_2}$ for ion and atom, respectively,
and by the angular-momentum quantum numbers $\ell$ and $m_\ell$ of the
relative motion of the atom and ion center of masses.  We label the
internal quantum numbers in the asymptotic basis by $\alpha = \{f_1 f_2
m_{f_1} m_{f_2} \}$, hence the asymptotic channels are characterized
by $i = |\alpha \ell m_\ell\rangle$.  At short distances it is more
convenient to characterize the channel states in terms of the total
electron spin $\mbf{S} = \mbf{s}_1 + \mbf{s}_2$ and total nuclear spin
$\mbf{I} = \mbf{i}_1 + \mbf{i}_2$, where $\mbf{s}_1$, $\mbf{s}_2$ are
electron spin of ion and atom respectively, and $\mbf{i}_1$, $\mbf{i}_2$
denote their nuclear spins, respectively. This basis is characterized
by two additional quantum numbers, the total hyperfine angular momentum
$\mbf{F}=\mbf{f}_1+\mbf{f}_2=\mbf{I}+\mbf{S}$ and its projection $M_f$
on the axis of quantization. We denote the internal quantum numbers in the
molecular basis by $\beta = \{ I S F M_F \}$, so the molecular channel
states are characterized by $j = |\beta \ell m_\ell \rangle$. The frame
transformation is a unitary transformation between channels $\alpha$
and $\beta$, that can be written as
\begin{align}
\label{U}
U_{\alpha\beta} & = ( f_1 f_2 m_{f1} m_{f2}|
I S F M_F)  = \sum_F (m_{f_1} m_{f_2}|F M_F)(f_1 f_2|I S).
\end{align}
Here, for simplicity we have omitted the quantum numbers $\ell$ and
$m_\ell$ that are conserved if one neglects weak dipolar
interactions. Symbol $(m_{f1}m_{f2}|F M_F)$ stands for the usual
Clebsch-Gordan coefficient
\begin{align}
(m_{f1}m_{f2}|F M_F)= & (-1)^{f_1-f_2+M_F} \sqrt{2F+1} \left(
\begin{array}{ccc}
f_1 & f_2 & F \\
m_{f_1} & m_{f_2} & M_{F}
\end{array}
\right),
\end{align}
while
\begin{align}
(f_1 f_2|I S) =& \sqrt{(2f_1+1)(f_2+1)(2I+1)(2S+1)}  \left\{
\begin{array}{ccc}
i_1 & s_1 & f_1 \\
i_2 & s_2 & f_2 \\
I & S & F
\end{array}
\right\},
\end{align}
is the transformation between $f_1 f_2$ and $IS$ coupling schemes, given in terms of the Wigner $9j$ symbol.

The knowledge of the analytical solutions for the polarization potential
suggest that the reference potentials can be chosen to contain only the
diagonal long-range part of the interaction matrix:

\begin{equation}
\label{Vdef}
V_{i}(r) = E_i^{\infty} +\frac{\hbar^2 \ell_i(\ell_i+1)}{2 \mu r^2} - \frac{C_{4}}{r^4} .
\end{equation}

According to Eqs.~\eqref{fhatS} and \eqref{ghatS} the wave functions
in pure polarization potential are singular at $r \rightarrow 0$. In
this case the standard boundary conditions $\mbf{F}(r) \rightarrow 0$
($r \rightarrow 0$) that is imposed on the physically meaningful
solution, has to be replaced by the boundary condition \eqref{fhatS}
with some short range phase $\phi_i$.

\subsection{Angular-momentum-insensitive quantum-defect matrix}

The exchange interaction that mixes asymptotic channel states $\alpha$
takes place usually in some range of distances $R_0$ of the order of few
tens of $a_0$. At such distances the interaction potential between species
is much larger than the hyperfine or Zeeman splittings in the presence
of an external magnetic field. Therefore it is convenient to define
the matrix $\mbf{Y}$ at distances $r \gtrsim R_0$, where the exchange
interaction is negligible, and $W_{ij}(r) \cong \delta_{ij} C_4/r^4$  for
$R_0 \lesssim r \ll R^\ast$. This approximation relies on the fact that
for $r \ll R^\ast$ one can safely neglect the centrifugal potential, the
asymptotic kinetic energy and hyperfine splittings. On the other hand, it
also ignores the higher order dispersion terms $C_6/r^6,C_8/r^8,\cdots$,
which only give relatively small corrections for the specific atom-ion
system considered in this paper \cite{Idziaszek2009}. Hence, for $R_0
\lesssim r \ll R^\ast$ the single-channel wave function is given by the
linear combination of the two WKB solutions \eqref{fghatS}.

Based on the arbitrariness of the reference potentials, we can choose
values $\phi_i$ at our convenience. For the following choice of short
range phases
\begin{equation}
\label{phi0}
\phi_i=0,
\end{equation}
the quantum-defect matrix takes a particularly simple form in the molecular ($IS$) basis
\begin{equation}
Y^{(IS)}_{\beta \beta^\prime} = \delta_{\beta \beta^\prime}[a_{S(\beta)}]^{-1},
\end{equation}
where $S(\beta)=0,1$ denotes the total electron spin in the channel
$\beta$, and $a_0 = a_s$ and $a_1 = a_t$ are the singlet and triplet
$s$-wave scattering lengths, respectively. By using the unitary
transformation \eqref{U} we find the quantum-defect matrix $\mbf{Y}$
in the basis of the asymptotic channel states
\begin{equation}
\mbf{Y}= \mbf{U} \mbf{Y}^{(IS)} \mbf{U}^\dagger.
\end{equation}

In the presence of an external magnetic field $B$, the total
transformation matrix is a product of $\mbf{U}$ and an additional unitary
matrix $\mbf{Z}(B)$ relating the bare ($B=0$) and dressed ($B\neq 0$)
channel states $|\Psi_\beta(B)\rangle$:
\begin{equation}
\label{BDressStates}
\left|\Psi_\alpha(B)\right\rangle = \sum_{\alpha^\prime} Z_{\alpha \alpha^\prime}(B) \left|\Psi_{\alpha^\prime}(B=0)\right\rangle.
\end{equation}
In this case the quantum-defect matrix reads
\begin{equation}
\mbf{Y}= \mbf{Z}(B) \mbf{U} \mbf{Y}^{(IS)} \mbf{U}^\dagger \mbf{Z}^\dagger(B).
\end{equation}
Apart from the shifts of the internal hyperfine states of the atom
and ion, the magnetic field in general also affects the motion of
the ion as a charged particle. This leads the Landau quantization
of the ion motion \cite{Landau} and can result in several interesting
scattering effects resembling the collisions in the quasi-1D confinements
\cite{Andrea2010}. There are, however, two limits when the ion cyclotron
motion can be neglected. First, when the characteristic size of the
Landau orbit $a_L = \sqrt{\hbar/m_i \Omega}$ with $\Omega=e B/m_i$
and $e$ denoting the ion charge, is much smaller than $R^\ast$. Second,
in the case when the ion is confined in the RF trap, which is typically
much tighter than the size of the Landau states. This second case is
however beyond the scope of the present analysis, and in the following
we assume the former condition to apply.

\subsection{Corrections to quantum-defect matrix for higher partial waves}
\label{Sec:SemiCorr}

According to the arguments presented in section~\ref{Sec:Frame}, at
distances $R_0 \ll R^{\ast}$ one can expect the dependence of the
quantum-defect matrix $\mbf{Y}$ on $\ell$ to be negligeable. However, as we will show later
with the example of a Na atom and a Ca$^+$ ion, this approximation works
well only for the few lowest partial waves. Already for $\ell \gtrsim
4$ one starts to observe some deviations of the quantum-defect parameter from
the exact numerical solutions for a realistic potential. In order to improve the
accuracy of the quantum-defect model we introduce a correction of
the short-range phase $\phi$ for nonzero relative angular momenta
$\ell$. Such correction can be obtained within the semiclassical theory, since the
modification of $\phi$ arises at distances where the real
potential $W_{ii}(r)$ differs from $V_i(r)$, a region well described in the
semiclassical approximation. We start from the semiclassical formula for
the radial wave function $\widetilde{f}_i(r)$ calculated in potential
$W_{ii}(r)$
\begin{equation}
\label{fsem}
\widetilde{f}_{i}(r) \cong \widetilde{k}_i(r)^{-1/2} \sin \widetilde{\beta}_i (r),
\end{equation}
where $\widetilde{k}_i(r) = \sqrt{2 \mu (E_i^{\infty} - W_{ii}(R)}/\hbar$
is the local wavevector, $\widetilde{\beta}_i(r) = \frac{\pi}{4} +
\int^r_{r_T}\!\mrm{d}x\,\widetilde{k}_i(x)$ is the WKB phase, and $r_T$
is the classical turning point. Here, we neglect the contribution from
the asymptotic kinetic energy, taking $E =E_i^{\infty}$. Eq.~\eqref{fsem}
can be rewritten in the following way
\begin{equation}
\label{fsem1}
\widetilde{f}_{i}(r) \cong \widetilde{k}_i(r)^{-1/2} \sin \left[ \beta_i(r)
+ \Delta_i(r) \right],
\end{equation}
where
\begin{equation}
\Delta_{i}(r) = \widetilde{\beta}_{i} (r) - \beta_i(r),
\end{equation}
$k_i(r) = \sqrt{2 \mu (E_i^{\infty}- V_i(R)}/\hbar$ and $\beta_i(r)
= \frac{\pi}{4} + \int^r_{r_T}\!\mrm{d}x\,k_i(x)$. At sufficiently
large distances, where the potential takes its asymptotic form
\eqref{Wasympt}, but still within the semiclassical regime $r \ll
R^{\ast}$, we require that the phases of $\widetilde{f}_{i}(r)$ and
of the solution $\hat{f}_{i}(r)$ calculated for $V_i(r)$ be equal. At
large $r$, $\Delta_{i}(r) \rightarrow \Delta_{i}(\infty) = const$,
and this happens already within the semiclassical domain $r \ll
R^{\ast}$, since the potentials $W_{ii}(r)$ and $V_{i}(r)$ have the same
long-range asymptotics. The $\ell$-dependent shift can be obtained
by comparing the $\Delta_{i}(\infty)$ computed for $\ell=0$ and $\ell >0$
\begin{equation}
\label{PhiCorr}
\delta \phi_i(\ell) = \left.\Delta_{i}(\infty)\right|_{\ell} - \left. \Delta_{i}(\infty) \right|_{\ell=0}.
\end{equation}
In this formula, $\left.\Delta_{i}\right|_{\ell}$ is calculated for
the actual value of $\ell$, while $\left. \Delta_{i} \right|_{\ell=0}$
represents a shift that can be incorporated into the definition of
the short-range phase $\phi$. By treating the centrifugal barrier in the
expressions for $k_i(r)$ and $\widetilde{k}_i(r)$ as a perturbation to
the interaction potential, we find a particularly simple result containing partial-wave quantum number $\ell$ only in the prefactor of an integral
\begin{equation}
\delta \phi_i(\ell) = \frac{\ell_i(\ell_i+1) \hbar}{2\sqrt{2 \mu}}
\int^\infty_{r_T}\!\frac{\mrm{d}x}{x^2} \left(
\frac{1}{\sqrt{C_4/x^4}} - \frac{1}{\sqrt{U_i(r)}},
\right)
\end{equation}
where $U_i (r) = W_{ii}(r) - E_i^{\infty} -\ell_i(\ell_i+1)/(2 \mu r^2)$.


\section{Radiative charge transfer}

In this section we develop a quantum mechanical description of the
radiative charge exchange process ${\rm A} + {\rm B}^+ \to {\rm A}^+ +
{\rm B} + h \nu$ with emission of a photon of frequency $\nu$. Since the
probability of charge exchange in heteronuclear atom-ion collisions is
small, the quantum transition rates can be described in the distorted wave
Born approximation (DWBA) \cite{taylor} treating the molecular dynamics
exactly and the interaction with the radiation field as a perturbation. If
weak dipolar interactions are neglected, the molecular dynamics conserves
the quantum numbers $\ell$ and $m_\ell$ of the orbital angular momentum
of the atomic fragments both in the initial and final molecular states.
Radiative transitions are induced to first order by the electric dipole
coupling, provided that the variation $|\Delta \ell|=1$.

We ignore at first nuclear spins, such that the collision
only involves two coupled molecular channels. After angular integrals
have been performed analytically using standard angular momentum
techniques, the radiative charge-exchange decay rate reads
(see, e.g., Ref.~\cite{zygel}):
\bea
\label{AEquant}
A(E) &=& \frac{ 64 \pi^4 }{3h c^3} \Big\{ \sum_{\ell}
\int \ud \left( h \nu \right) \nu^3 \left[
\left( \ell+1 \right)
| \langle E^{\prime} \ell+1 |d(r)| E \ell  \rangle |^2
+ \ell  |  \langle E^{\prime} \ell-1 |d(r)| E \ell  \rangle  |^2
\right]
\nonumber \\
& & + \sum_{v^\prime } \nu_{v^\prime }^3 \left[   \left( \ell+1 \right)
 | \langle v^{\prime} \ell+1 |d(r)| E \ell  \rangle |^2
+ \ell  |  \langle v^{\prime} \ell-1 |d(r)| E \ell  \rangle  |^2
\right] \Big\}.
 \eea
This equation is expressed in terms of reduced dipole matrix elements
for free-free $\langle E^{\prime} \ell^{\prime} |d(r)| E \ell  \rangle $
transitions, in which the atom and the ion remain unbound, and free-bound
$\langle v^{\prime} \ell^{\prime} |d(r)| E \ell  \rangle $ transitions,
in which the colliding pair forms a molecular ion.

The $|E \ell \rangle$ scattering state is energy-normalized with
incoming wave boundary conditions in the entrance potential of ${\rm
A}  {\rm B}^+$ state. The primed quantities $|E^\prime \ell^\prime
\rangle$ and $|v^\prime \ell^\prime \rangle$ represent respectively
energy-normalized scattering states with outgoing wave boundary conditions
and unit normalized bound states of the ${\rm A}^+ {\rm B}$ molecular
complex. Energy conservation requires $h \nu = E-E^\prime+\delta$ and
$h \nu_{v^\prime}=E-E_{v^\prime}+\delta$ for free-free and free-bound
transitions, where $\delta$ is the difference of ion and atom ionization
potentials.

While this approach is physically transparent and has the advantage of
providing detailed information on the products of the charge-exchange
process, the generation of dipole moments as a function of photon
energy and of free-bound matrix elements for all ro-vibrational levels is
computationally costly.
A significantly simpler approach discussed in~\cite{Tellinghuisen} approximates
the sum over all continuum and bound transitions with a simple average
of a space-varying Einstein coefficient for spontaneous emission over
the initial scattering wavefunction
\be
A(E)=\sum_{\ell} (2\ell +1) \langle E \ell | \bar{A}(r) | E \ell \rangle,
\label{closure}
\ee
with
\be
\label{Abar}
\bar{A}(r)=\frac{64 \pi^4 \nu^3(r) d^2(r)}{3hc^3}.
\ee

Eq.~(\ref{closure}) is inspired by the exact quantum-mechanical closure
relation over scattering and bound states in the final channel, and by the
classical Frank-Condon principle relating the potential energy difference
with the energy of a photon emitted at interatomic separation $r$, $h
\nu(r) = W_i (r)-W_f(r)$ ~\cite{Tellinghuisen}. In the next subsection we
cast this equation in a particularly expressive form using a combination
of semiclassical and MCQDT approaches.

\subsection{Semiclassical model}
\label{Sec:SemModel}

The radiative charge transfer takes places at relatively short distances,
where the electronic wave functions of the atom and ion start to overlap
and the dipole matrix element is non negligible. At such distances one
can calculate the matrix element of $\bar{A}(r)$ in the semiclassical
approximation
\begin{equation}
\label{Amatr}
\langle E \ell|\bar{A}(r)| E \ell \rangle \cong \frac{2}{h} C^{-2}(E,\ell) \int_{R_\mrm{min}}^\infty dr \frac{\bar{A}(r)}{v(r)}.
\end{equation}
Here, $R_\mrm{min}$ denotes the classical turning point in the potential $U(r)$ of the entrance channel, and
\begin{equation}
v(r) = \sqrt{\frac{2}{\mu}} \sqrt{E-U(r)-\frac{ \hbar^2 (\ell+\frac12)^2}{2 \mu r^2}}
\end{equation}
is the classical velocity of a particle in the entrance channel at
the distance $r$. The MQDT function $C^{-2}(E,\ell)$ provides the
proper scaling of the semiclassical wave function at short distances,
with respect to its long-range asymptotic behavior, given by the
energy-normalized functions $|E \ell \rangle$. The total cross section
for the charge exchange can be calculated from \cite{Julienne}
\begin{equation}
\label{SigmaTr}
\sigma_\mrm{tr} = \frac{2 \pi}{k^2} h A(E).
\end{equation}

In the regime of ultracold energies one can neglect the contribution
of the energy $E$ to $v(r)$ at distances where the radiative charge
transfer occurs. Similarly, one can also omit the contribution from
the centrifugal barrier. In this approximation $v(r) \approx \sqrt{-2
U(r)/\mu}$, and the energy and angular momentum dependence enters only
through the MQDT function $C^{-2}(E,\ell)$
\begin{equation}
\label{AmatrAppr}
\langle E \ell| \bar{A}(r)| E \ell \rangle \approx \frac{1}{h} C^{-2}(E,\ell) P_\mrm{tr}
\end{equation}
Here, $P_\mrm{tr} = 2 \int_{R_\mrm{min}}^\infty dr\, \bar{A}(r)/v(r)$
is the probability of the photon emission during a single collision,
which in our approximation is a constant. This way the cross section
for the charge transfer event $\sigma_\mrm{tr}$ can be written as
\begin{equation}
\label{SigmaTr2}
\sigma_\mrm{tr}(E) = \frac{2 \pi}{k^2} P_\mrm{tr} \sum_{\ell} (2 \ell + 1) C^{-2}(E,\ell) .
\end{equation}

It is instructive to investigate the classical, high energy limit of
the charge transfer rate, where we expect the Langevin theory to be
applicable. Starting from Eq.~\eqref{closure} with the semiclassical
approximation \eqref{Amatr}, then setting $C^{-2}(E,\ell) \approx 1$ at
$E \gg E^\ast$, and replacing summation over $\ell$ by an integration over
impact parameter $b= (\ell + \frac{1}{2})/k$, we obtain \cite{Makarov2003}
\begin{equation}
\label{sigmacl}
\sigma_\mrm{tr}^\mrm{class}(E) = 2 \pi \sqrt{2 \mu} \int_0^\infty\!\!\!\!db\,b\! \int_{R_\mrm{min}}^\infty\!\!\!\!dr \frac{\bar{A}(r)}{\sqrt{E-U(r)-E b^2/r^2}},
\end{equation}
Eq.~\eqref{sigmacl} can be further simplified, first by replacing the
inner integral by a constant probability $P_\mrm{tr}$, and then by
performing the remaining integral over $b$. We perform the integration
over $b$ to some maximal value of the impact parameter $b_\mrm{max}$,
given by the height of the centrifugal barrier $E_\mrm{max}(\ell) =
\frac{1}{4} E^\ast \ell^2(\ell+1)^2$. Thus, the particle can penetrate
the inner part of the potential only for $E > E_\mrm{max}$, which imposes
the upper bound on $b$: $b_\mrm{max} = \sqrt{2 R^\ast/k}$. This yields
for the cross section
\begin{equation}
\label{sigmacl1}
\sigma_\mrm{tr}^\mrm{class}(E) = \frac{2 \pi R^\ast}{k} P_\mrm{tr},
\end{equation}
and for the charge transfer collision rate $K_\mrm{tr} = \sigma_\mrm{tr}(E) v.$
\begin{equation}
\label{Kcl1}
K_\mrm{tr}^\mrm{class} = \frac{h R^\ast}{\mu} P_\mrm{tr} = 2 \pi \sqrt{\frac{2C_4}{\mu}} P_\mrm{tr}.
\end{equation}
At large energies the charge transfer rate acquires a constant value. The
latter formula agrees with predictions of the Langevin theory, assuming
that all the classical trajectories that fall down on the scattering
center lead to a reaction \cite{Vogt1954}. In our case the probability
of a reaction in a single collision event is given by $P_\mrm{tr}$,
whereas in the case of homonuclear collisions, where the resonant charge
transfer takes place $P_\mrm{tr} = \frac12$ \cite{Grier2009}.

It is convenient to express the charge transfer rate in the quantum regime in terms of $K_\mrm{tr}^\mrm{class}$:
\begin{align}
\label{Ktr}
K_\mrm{tr}(E) & = K_\mrm{tr}^\mrm{class} Q(E) \\
\label{Qdef}
Q(E) & = \frac{1}{2 k R^\ast} \sum_{\ell} (2 \ell + 1) C^{-2}(E,\ell)
\end{align}
In this form the charge transfer rate is a product of a constant
classical rate, that depends on the particular atom-ion system, and a
quantum factor $Q(E)$ that is universal and depends only on the specific
atom-ion combination through the characteristic energy $E^\ast$ and
the short-range phase $\phi$. In this way the whole energy dependence
will be given by $Q(E)$. In the limit of high energy ($E \to \infty$)
$Q(E) \to 1$. In the limit of small energies ($E \to 0$), only $s$
wave contributes and $Q(E) \to \frac{1}{2} (1 + \cot^2 \phi)$.

\begin{figure}
\begin{indented}
\item[]\includegraphics[width=8.6cm,clip]{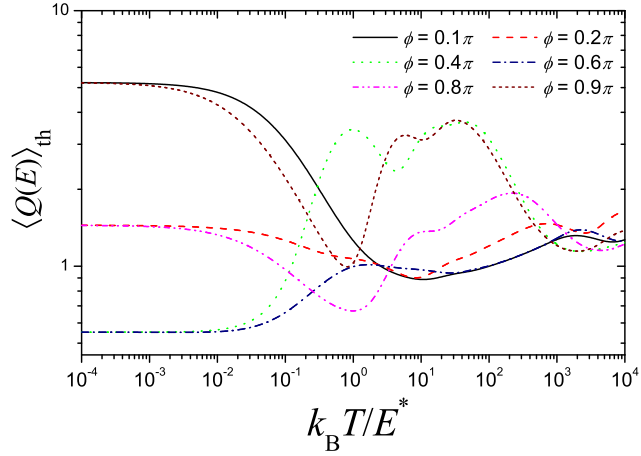}
\end{indented}
\caption{Thermally averaged quantum contribution $\langle Q(E) \rangle_\mrm{th}$ to the charge transfer rate versus temperature for selected values of the short-range phase.
\label{Fig:CERates}
}
\end{figure}

In Fig.~\ref{Fig:CERates} we show $Q(E)$ averaged over thermal distribution
\begin{equation}
\langle Q(E) \rangle_\mrm{th} \equiv \frac{2}{\sqrt{\pi} (k_B T)^{3/2}}
\int_0^{\infty}\!\!dE\,  Q(E) \sqrt{E} e^{-E/k_B T}
\end{equation}
for selected values of the short range phase. We observe that even at
large energies $E \sim 10^4 E^\ast$, $\langle Q(E) \rangle \neq 1$,
due to the contribution of shape resonances. The detailed structure of
resonances, however, is washed out by the thermal average.

\section{Results for $^{40}\mrm{Ca}^{+}$ and $^{23}\mrm{Na}$}

We now apply our quantum-defect model to $^{40}\mrm{Ca}^{+}$ and
$^{23}\mrm{Na}$, a system whose {\it ab initio} potential energy curves
are known \cite{Makarov2003}. This allows us to compare the predictions
of our analytical approach with the full solution obtained by solving
numerically the coupled-channel Schroedinger equation \eqref{RadialSchr}
with potentials of Ref. \cite{Makarov2003}.

\subsection{Channel states}

\begin{figure}
\begin{indented}
\item[]\includegraphics[width=8.6cm,clip]{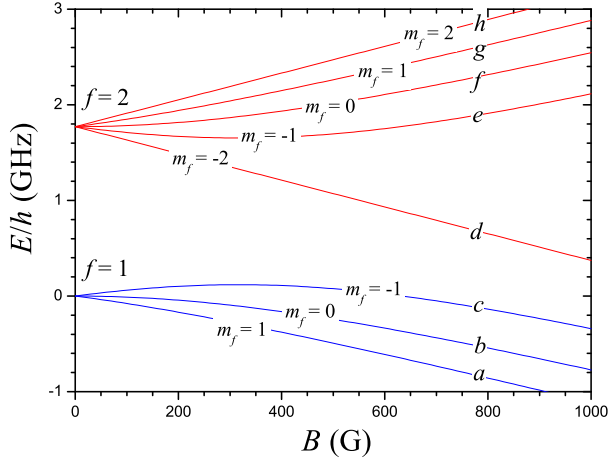}
\end{indented}
\caption{
\label{Fig:Hyperfine}
Hyperfine structure of $^{23}\mrm{Na}$. Zeeman sublevels versus magnetic field $B$ are shown.}
\end{figure}

The $^{40}\mrm{Ca}$ ion has vanishing nuclear spin ($i_1 = 0$, $s_1 =
1/2$). We label its ground-state sublevels in an external magnetic
field in increasing energy $|a_1\rangle \equiv |f=1/2,m_f=-1/2\rangle$
and $|b_1\rangle \equiv |f=1/2,m_f=1/2\rangle$. The $^{23}\mrm{Na}$
has nuclear spin $i_2 = 3/2$, hence its hyperfine angular momentum
can take the values $f=1$ and $2$. The hyperfine structure of $^{23}\mrm{Na}$
is shown in Fig.~\ref{Fig:Hyperfine}. The figure adopts the
standard $|a_2\rangle $, $|b_2\rangle, \cdots$ notation to label states
in increasing energy in the magnetic field, where in the weak field limit one has the identification $|a_2\rangle \equiv
|f=1,m_f=1\rangle$, $|b_2\rangle \equiv |f=1,m_f=0\rangle, \cdots$. In the
presence of an external magnetic field, only the projection $M_J$ of the
total angular momentum $\mbf{J} = \mbf{f}_1 + \mbf{f}_2 + \mbf{l}$ is
conserved during a collision. However, if we ignore small anisotropic spin-spin interactions
giving rise to the long-range dipole-dipole force, the states with
different angular momentum $\mbf{l}$ will not be coupled, and both $\ell$
and $m_\ell$ will be conserved. Hence, $M_F$ will also be conserved and
we can restrict our discussion to subspaces of constant
$M_F$. In most of our calculations we will consider collisions within
the subblock of $M_F=1/2$, which contains the lowest energy channel state
$| a_1 a_2\rangle$.  For $M_F=1/2$  there are four possible scattering
channels, listed in Table~\ref{Tab:ChFM} together with their threshold
energies at $B=0$. Table~\ref{Tab:ChIS} lists the channel states in the
molecular basis.

\Table{\label{Tab:ChFM} Scattering channels of $^{40}\mrm{Ca}^{+}$ and $^{23}\mrm{Na}$ for $M_F=1/2$ in the asymptotic representation.}
\br
& $\alpha$ & $|f_1,m_{f_1},f_2,m_{f_2}\rangle$ & $E_\alpha^{\infty}/h$(GHz)\\
\mr
1 & $|a_1a_2\rangle$ & $|\frac12,-\frac12,1,1\rangle$ & 0\\
2 & $|b_1b_2\rangle$ & $|\frac12,\phantom{-}\frac12,1,0\rangle$ & 0\\
3 & $|a_1g_2\rangle$ & $|\frac12,-\frac12,2,1\rangle$ & 1.77163\\
4 & $|b_1f_2\rangle$ & $|\frac12,\phantom{-}\frac12,2,0\rangle$ & 1.77163 \\
\br
\endTable
\Table{\label{Tab:ChIS} Scattering channels of $^{40}\mrm{Ca}^{+}$ and $^{23}\mrm{Na}$ for $M_F=1/2$ in the molecular $\{IS\}$ representation.}
\br
 & $\beta$ & $|F,M_{F},I,S\rangle$ \\
\mr
1 & $\frac{1}{2\sqrt{2}}|a_1a_2\rangle+\frac12|b_1b_2\rangle-\sqrt{\frac38}|a_1g_2\rangle+\frac12|b_1f_2\rangle$
& $|\frac32,\frac12,\frac32,0\rangle$ \\
2 & $\sqrt{\frac23}|a_1a_2\rangle-\frac{1}{\sqrt{3}}|b_1b_2\rangle$
& $|\frac12,\frac12,\frac32,1\rangle$ \\
3 & 
$\sqrt{\frac{5}{24}}|a_1a_2\rangle+\sqrt{\frac{5}{12}}|b_1b_2\rangle-\frac{3}{\sqrt{40}}|a_1g_2\rangle-\sqrt{\frac{3}{20}}|b_1f_2\rangle$ & $|\frac32,\frac12,\frac32,1\rangle$ \\
4 & $\sqrt{\frac25}|a_1g_2\rangle+\sqrt{\frac35}|b_1f_2\rangle$ & $|\frac52,\frac12,\frac32,1\rangle$ \\
\br
\endTable

The quantum-defect matrix $\mbf{Y}$ in the molecular $\{IS\}$ basis is
diagonal and for the assumed channel numbering it reads
\begin{equation}
\mbf{Y}^\mrm{(IS)} = \left( \begin{array}{cccc}
(a_s)^{-1} & 0 & 0 & 0 \\
0 & (a_t)^{-1} & 0 & 0 \\
0 & 0 &(a_t)^{-1} & 0  \\
0 & 0 & 0 & (a_t)^{-1} \\
\end{array}\right)
\end{equation}
Using the frame transformation \eqref{U} one can easily find $\mbf{Y}$ in the asymptotic channel representation
\begin{equation}
\label{Yfmf}
\mbf{Y} =
\left(
\begin{array}{cccc}
\dsp{\frac{1}{a_t}+\frac{1}{8 a_c}}& \dsp{\frac{\sqrt{2}}{8 a_c}} & \dsp{-\frac{\sqrt{3}}{8 a_c}} & \dsp{\frac{\sqrt{2}}{8 a_c}} \\
\dsp{\frac{\sqrt{2}}{8 a_c}} & \dsp{\frac{1}{a_t}+\frac{1}{4 a_c}} & \dsp{-\frac{\sqrt{6}}{8 a_c}} & \dsp{\frac{1}{4 a_c}} \\
\dsp{-\frac{\sqrt{3}}{8 a_c}} & \dsp{-\frac{\sqrt{6}}{8 a_c}} & \dsp{\frac{1}{a_t}+\frac{3}{8 a_c}} & \dsp{-\frac{\sqrt{6}}{8 a_c}} \\
\dsp{\frac{\sqrt{2}}{8 a_c}} & \dsp{\frac{1}{4 a_c}} & \dsp{-\frac{\sqrt{6}}{8 a_c}} & \dsp{\frac{1}{a_t}+\frac{1}{4 a_c}}
\end{array}
\right),
\end{equation}
where $1/a_c = 1/a_s - 1/a_t$ is the coupling parameter characterizing the
strength of coupling between channels. We note that for similar triplet
and singlet scattering lengths $1/a_c=0$, the channels are uncoupled,
and no interchannel resonances occur.

In the presence of a magnetic field, the energies of the hyperfine states of Na atom are given by the Breit-Rabi formula \cite{Breit1931}
\begin{equation}
\label{Ebr}
E_{fm_f}^\mrm{(Na)}(B) = \frac{E_\mrm{hf}}{2} + (-1)^{f-\frac12} \sqrt{\left(\frac{E_\mrm{hf}}{2} + x\right)^2 + y^2}
\end{equation}
with
\begin{align}
x=& \frac{g \mu_B B m_f}{1+2 i} \nonumber \\
y=& (-1)^{2(i+m_f)} g \mu_B B \frac{\sqrt{\left(i+\tst{\frac12}\right)^2-m_f^2}}{1+2i}, \nonumber
\end{align}
where $g$ is Land\'e factor, and $\mu_B$ is the Bohr magneton. By
including in the Hamiltonian additional small couplings of the
magnetic field to the nuclear spin, $H(B) = H(0) + \mu_B \mbf{B}
(g_\mrm{j} \mbf{j} + g_\mrm{i} \mbf{i})$, the effective $g$-factor is
given by $g = g_\mrm{j} - g_\mrm{i}$, and the levels acquire an additional
$B$-dependent shift $\Delta E_{fm_f}^\mrm{(Na)}(B) =g_\mrm{i} \mu_B B
m_\mrm{f}$. Here, $g_\mrm{j}$ and $g_\mrm{i}$ are Land\'e factors for
the total orbital angular momentum $\mbf{j}= \mbf{L} + \mbf{s}$ and the
nuclear spin $\mbf{i}$, respectively. In Ca ion, which has no hyperfine
structure, the levels shift according to the standard Zeeman formula
\begin{equation}
\label{Ezeem}
E_{f m_f}^\mrm{(Ca)}(B) = \mu_B B g_{j} m_{f}
\end{equation}

The transformation matrix $\mbf{Z}$ from the bare ($B=0$) to the dressed
channel states ($B\neq0$) can be easily found from the transformation
matrices $\bar{\mbf{Z}}^{(k)}$, connecting bare and dressed states of ion
($k=1$) and atom ($k=2$). For the ion we have simply $\bar{\mbf{Z}}^{(1)}
= \mbf{1}$, while for the atom
\begin{equation}
\bar{\mbf{Z}}^{(2)} = \frac{1}{w}\left(
\begin{array}{cc}
x+E_{\bar{f}_2 m_\mrm{f2}}^\mrm{(Na)}(B) &
-y \\
y &
x+E_{\bar{f}_2 m_\mrm{f2}}^\mrm{(Na)}(B)
\end{array}
\right)
\end{equation}
where $\bar{f}_2 = i_2 + 1/2=2$, and $w= \sqrt{y^2 + (x+E_{\bar{f}_2 m_\mrm{f2}})^2}$. Then the total transformation matrix $\mbf{Z}$ for the assumed channel numbering reads
\begin{equation}
\mbf{Z} = \left( \begin{array}{cccc}
\bar{Z}^{(2)}_{11} & 0 & \bar{Z}^{(2)}_{12} & 0 \\
0 & \bar{Z}^{(2)}_{11} & 0 & \bar{Z}^{(2)}_{12} \\
\bar{Z}^{(2)}_{21} & 0 &\bar{Z}^{(2)}_{22} & 0  \\
0 & \bar{Z}^{(2)}_{21} & 0 & \bar{Z}^{(2)}_{22} \\
\end{array}\right)
\end{equation}

\subsection{Bound states}

\begin{figure}
\begin{indented}
\item[]\includegraphics[width=8.6cm,clip]{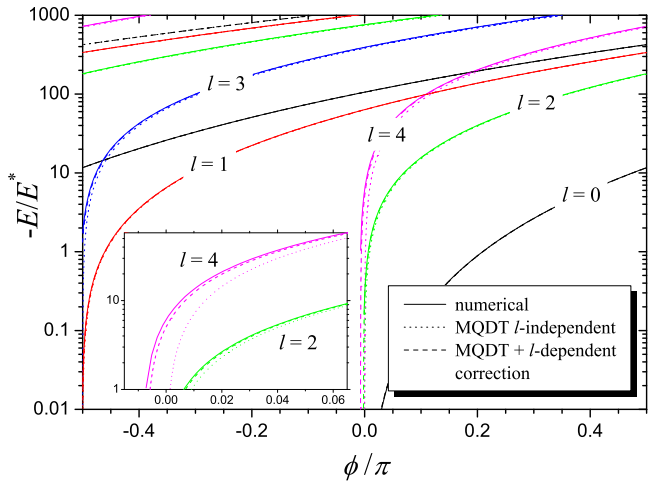}
\end{indented}
\caption{
\label{Fig:Eb}
Energies of the bound states versus the short-range phase $\phi$ in
the triplet channel $a^3 \Sigma^+$ for few lowest partial waves. The
numerical results obtained by solving the Schr\"odinger equation with
{\it ab-initio} potential of Ref.~\cite{Makarov2003} (solid lines),
are compared with predictions of MQDT assuming angular-momentum
insensitive short-range phase (dotted lines), and with MQDT including
the $\ell$-dependent correction \eqref{PhiCorr} to the short-range phase
(dashed lines). The inset zooms in the region around $\phi = 0$.}
\end{figure}

In this section we discuss properties of bound states in a single
scattering channel. Fig.~\ref{Fig:Eb} shows a sample bound state spectrum
for $a^3 \Sigma^+$ molecular potential. The numerical calculations based
on the {\it ab initio}  Na-Ca$^+$ potentials of Ref.~\cite{Makarov2003}
with the quantum-defect theory with and without the $\ell$-dependent
correction \eqref{PhiCorr} to the short-range phase are compared. For
simplicity in the numerical calculation we neglect the higher-order
dispersion terms in the potential, so to isolate the effect of the
centrifugal barrier, in particular for high order partial waves. We note
that the quantum-defect model assuming the same short-range phase $\phi$
for all partial waves starts to deviate already for $\ell=3$, whereas the
inclusion of the correction \eqref{PhiCorr} greatly improves the agreement
between MQDT model and numerical solution. The inset shows magnification
of the region around $\phi=0$. For $\ell=4$ still some small discrepancy
between the numerical and corrected MQDT results can be observed. Tha
latter may originate from effects going beyond the WKB approximation, which
was assumed to hold in the derivation of Eq.~\eqref{PhiCorr}.

\subsection{Radiative charge transfer}

In the case of Na-Ca$^+$ collisions the charge transfer process
occurs due to transitions from a singlet $A^1 \Sigma$ channel to
the absolute ground molecular term ${\rm Na}  {\rm Ca}^+(A^1 \Sigma)
\to {\rm Na}^+ {\rm Ca}(X^1 \Sigma) $. For its description we first
use the DWBA leading to Eq.~\eqref{AEquant}, which allows the
final states of the charge exchange process to be individually identified. In order
to calculate the free-free transition dipole elements a two coupled
channel model is setup, comprising $W_i(r)=V\left( A^1 \Sigma \right)$
and $W_f(r)=V\left( X^1 \Sigma \right)$ diagonal molecular potentials
with $\ell$ and $\ell^\prime$ centrifugal barriers, coupled by a $  \xi
d(r) $ term, with $\xi$ a parameter to be optimized. In a field-dressed
approach, a vertical energy shift is introduced so to guarantee total,
photons plus atoms, energy conservation. If the $\xi$ parameter is chosen
so to insure validity of first order perturbation theory, in the DWBA the
required reduced matrix element is simply proportional to the transition
scattering matrix element $s_\xi=s (E^\prime \ell^\prime \leftarrow E
\ell; \xi)$ obtained numerically
\be
   s_\xi
= -2\pi i \xi  \langle E^{\prime} \ell^\prime |d(r)| E \ell  \rangle .  \ee

\begin{figure}
\begin{indented}
\item[]\includegraphics[width=8.6cm,clip]{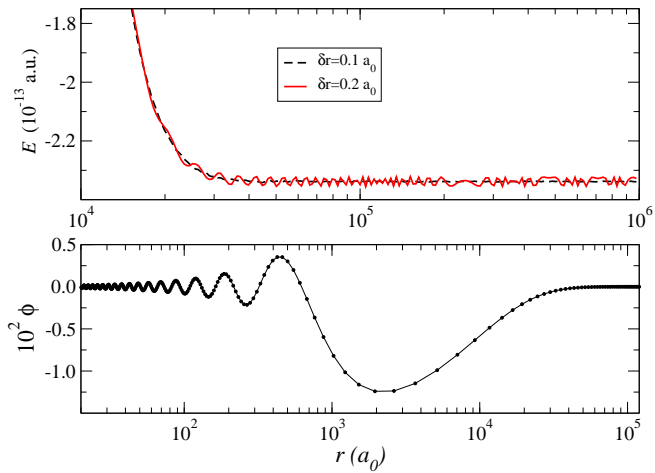}
\end{indented}
\caption{
\label{Fig:Conv}
The closest to threshold $\ell=0$ vibrational wavefunction for $s$-wave scattering
length $a=5 R^\ast$ (lower panel) showing the points of the adaptive numerical grid.
Convergence of the molecular energy is also shown (upper panel) as a function
of the numerical grid size for two values of the radial step $\delta r$ near the potential minimum (see text).
}
\end{figure}

The free-bound reduced dipole elements are obtained following a different
method. As a first step, the vibrational wavefunctions of the $X^1
\Sigma$ potential are generated for each angular momentum $\ell^\prime$
using a grid method. To this aim use of the step-adaptive approach
of Ref.~\cite{tiesinga} is essential. In fact, due to the long-range
character of the $r^{-4}$ interaction, numerical convergence of near
threshold levels is attained for a grid extending to large distances.
To keep the number of grid points reasonable ($\sim 5000$), the local
step size is made to vary over five orders of magnitude from the
potential well to the asymptotic region using an appropriate scaling
function ~\cite{tiesinga}.  Fig.~\ref{Fig:Conv} shows for instance the
last rotationless vibrational wave function for $a=5 R^\ast$, as well as
the convergence rate of the corresponding eigenenergy for two different
step sizes $\delta r$ of the numerical grid near the potential minimum.
Note that for this specific molecular level, convergence begins to be
attained for a box of $~10^5 a_0$ size.  The relative energy accuracy
is on the order of $10^{-2}$ for $\delta r=0.2 a_0$, and increases
by one order of magnitude for $\delta r=0.1 a_0$.
As a second step, the overlap $\langle v^{\prime} \ell^{\prime} |d(r)|
E \ell  \rangle  $ with the initial scattering wavefunction of angular
momentum $\ell$ is computed using a standard propagation code. The number
of needed $\ell$ values is determined by numerical convergence.

In the alternative approach based on the approximate sum rule
Eq.~\eqref{closure}, the expectation value $\langle E \ell | \bar{A}(r)
| E \ell \rangle $ are extracted numerically from the elastic scattering
matrix element $s_\xi=s (E \ell \leftarrow E \ell; \xi)$ for a single
channel potential $W_i$ with angular momentum $\ell$ perturbed by a
$\xi \bar{A}(r)$ term. The scaling parameter $\xi$ has to be
chosen so to be in the linear regime. With this proviso, letting $s_0$
be the scattering matrix for the unperturbed $W_i$ potential, the needed
matrix element is determined from the DWBA as
\be
s_0^* (s_\xi - s_0) = -2\pi i \xi \langle E \ell | \bar{A}(r) | E \ell \rangle .
\ee
The prefactor $s_0^*$ arises from the fact that the expectation value on
the right hand side has to be evaluated between scattering states with
outgoing boundary conditions whereas transition matrix elements in the
DWBA present incoming wave boundary conditions in the exit channel.

\begin{figure}
\begin{indented}
\item[]\includegraphics[width=8.6cm,clip]{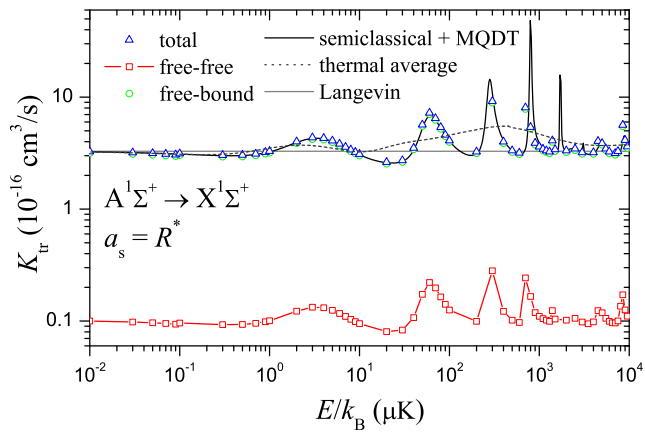}
\end{indented}
\caption{
\label{Fig:Rates}
Rates of the radiative charge transfer due to the transitions between
$A^{1} \Sigma^{+}$ and $X^{1} \Sigma^{+}$ states versus energy for the
single scattering length $a_s=R^\ast$. The full numerical calculation
based on the Fermi golden rule Eq.~\eqref{AEquant} (triangles) is compared
to the quantum-defect model assuming semiclassical description of the
charge transfer process (solid line). In addition we show the contribution
from free-free (squares) and free-bound (circles) transitions obtained
from the numerical calculation, the thermal average obtained from the
quantum-defect model (dashed line) and the charge transfer rate given
by the Langevin theory (gray solid line).
}
\end{figure}

The charge transfer rates we obtain are shown in Fig.~\ref{Fig:Rates}
as a function of the collision energy for a sample value of the singlet
scattering length, $a_s = R^\ast$. The figure compares the numerical
result calculated by summing contributions from all free-free and
free-bound transitions Eq.~\eqref{AEquant} with the MQDT model assuming
the semiclassical charge-transfer probability Eq.~\eqref{Ktr}. In
the former case we additionally plot separately contributions from
free-free and free-bound transitions. The MQDT calculation includes the
$\ell$-dependent correction \eqref{PhiCorr}. We observe that at energies
larger than $10\mu$K, the rate of charge exchange exhibits several peaks
due to the shape resonances. The MQDT model agrees well with the
full numerical calculations, except in the mK regime where high order
partial waves are contributing. In this range of energies the discrepancy
is due to the corrections to the short-range phase for large
$\ell$ that are beyond the applicability of the semiclassical formula
\eqref{PhiCorr}. Finally, the approximation based on Eq.~\eqref{closure}
is in full agreement with the numerically exact result and is not shown
in the figure.

\subsection{Population of vibrational states in the charge transfer process}

According to Fig.~\ref{Fig:Rates} the main outcome of the charge transfer
process are molecular ions. It is interesting to analyze the dependence
of the charge exchange rate on the vibrational quantum number of the
molecule, {\it i.e.} the vibrational distribution of the product molecular
ions. In the full quantum treatment based on the numerical calculation
of initial and final wave functions, such distribution is proportional
to the transition matrix elements in Eq.~\eqref{AEquant}. In the MQDT
approach it can be obtained using the semiclassical approximation, in
analogy with the derivation presented in Section~\ref{Sec:SemModel}. We
start from Eqs.~\eqref{closure} and \eqref{Amatr}, and obtain the
probability of the charge transfer process normalized per vibrational
quantum number $\vartheta$ in the final state
\begin{equation}
\label{dAdn}
\frac{dA}{d\vartheta}(E^\prime) = \frac{1}{\rho(E^\prime) \left|\frac{d U}{d r}(E^\prime)\right|} \frac{2}{h} \frac{A(r)}{v(r)} \sum_{J} (2 J + 1)  C^{-2}(E,J),
\end{equation}

Here, $\rho(E) = \frac{d \vartheta}{d E}$ is the density of states
in the exit channel and $V(r)  =   W_\mrm{i}(r) - W_\mrm{f}(r)$
is the difference between interaction potentials of the entrance
(initial) and the exit (final) channels. The distance $r$ at which all
$r$-dependent quantities are calculated is related to the final energy by
the Frank-Condon principle: $ E + \delta - E^\prime = \hbar \omega(r) =
V(r)$, with $\delta = E_\mrm{i}^{\infty} - E_\mrm{f}^{\infty}$ denoting
the difference of the dissociation energies of the entrance and exit
channels, respectively. The density of states $\rho(E)$ can be calculated
from the LeRoy-Bernstein formula \cite{LeRoy}. In the the case of $r^{-4}$
potential it yields
\begin{equation}
\rho(E) = \left|\frac{d \vartheta}{d E}\right| = \frac{1}{E^\ast} \frac{\Gamma(\frac34)}{2 \sqrt{\pi} \Gamma(\frac14)} \left(-\frac{E}{E^\ast}\right)^{-3/4}.
\end{equation}

\begin{figure}
\begin{indented}
\item[]\includegraphics[width=8.6cm,clip]{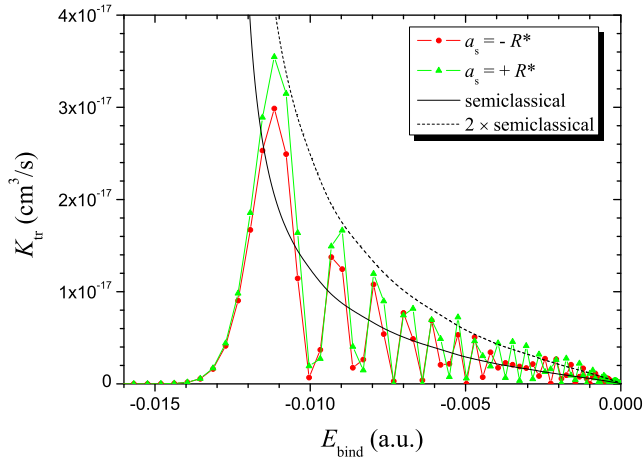}
\end{indented}
\caption{
\label{Fig:RatesDistr}
Rates of the radiative charge transfer due to the transitions between
$A^{1} \Sigma^{+}$ and $X^{1} \Sigma^{+}$ states versus energy of the
vibrational state in the exit channel, for two values of the singlet
scattering length $a_s$ (lines with symbols). The solid line shows the
semiclassical result obtained from Eq.~\eqref{dAdn}, while the dashed
line is the same result multiplied by two, which gives approximately
the amplitude of the quantum oscillations.}
\end{figure}

In Fig.~\eqref{Fig:RatesDistr} we show the charge transfer rate from
$A^{1} \Sigma^{+}$ channel to the bound states of the $X^{1} \Sigma^{+}$
channel, calculated numerically in the DWBA. The numerical points
are obtained for two values of the singlet scattering length: $a_s =
\pm R^\ast$. The solid line shows prediction of Eqs.~\eqref{dAdn},
with the total cross section related to the probability $A$ by
Eq.~\eqref{SigmaTr}. We note that the full quantum result predicts
oscillations of the distribution over final states, whereas the
semiclassical theory leads to a smooth behavior, which can be interpreted
as the distribution averaged over the quantum oscillations. The origin
of these oscillations in the quantum result is due to the
effects of the phase matching between the entrance and the exit channel
wave functions, which is not present in the semiclassical result assuming
Frank-Condon approximation.

\subsection{Magnetic Feshbach resonances at zero energy}

In this section we analyze magnetic Feshbach resonances in the limit
of zero energy when only $s$-wave scattering is present. We neglect the
effects of the magnetic field on the translational ion motion.
For the Na-Ca$^+$ system of interest, the radius of the Landau orbit
$a_L$ becomes equal to $R^\ast$ at $B_L =1086 G$. Hence, the presented
analysis is valid as long as $B \ll B_L$.

\begin{figure}
\begin{indented}
\item[]\includegraphics[width=8.6cm,clip]{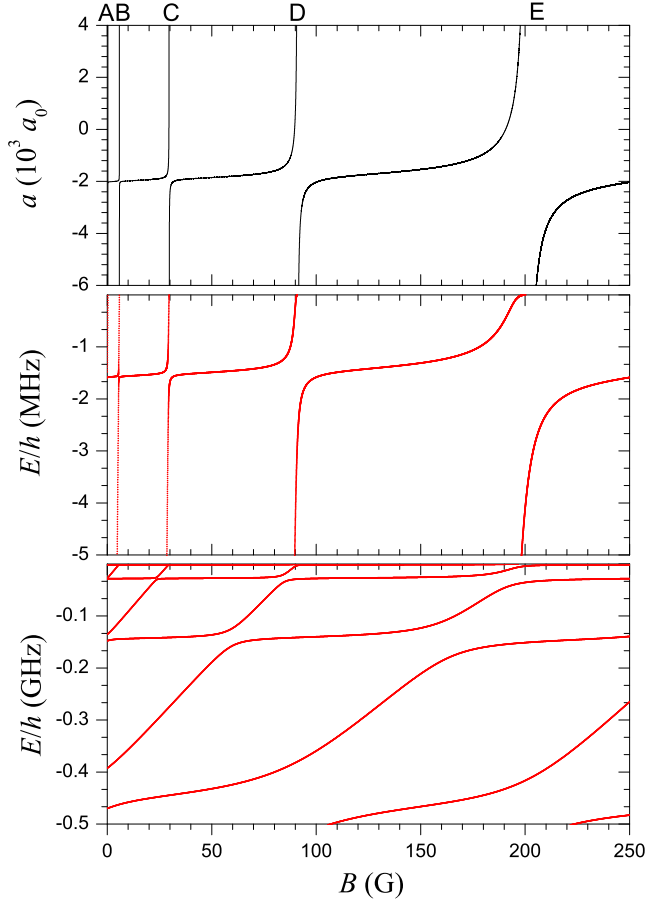}
\end{indented}
\caption{
\label{Fig:Feshbach0}
Scattering length (upper panel) and energies of the $s$-wave bound
states (middle and lower panel) versus the magnetic field strength $B$
for collisions of Na with $^{40}$Ca$^+$ calculated using MQDT model for
some typical singlet and triplet scattering lengths: $a_s =R^\ast=2081
a_0$ and $a_t =-R^\ast = -2081 a_0$, respectively. The capital letters
in the upper panel label the Feshbach resonances.}
\end{figure}

Figure~\ref{Fig:Feshbach0} shows the variation of the $s$-wave scattering
length versus magnetic field for two sample values of the singlet
and triplet scattering lengths: $a_s = R^\ast = 2081 a_0$ and $a_t =
-R^\ast = -2081 a_0$, respectively. The resonances can
be assigned to the particular bound state crossing the threshold of the
$\alpha=|a_1 a_2 \rangle$ channel, as one can verify by inspection of the
bound state energies in the bottom panel. The resonances are labeled
by capital letters from A to E. The middle panel shows a close-up
of the bound-state spectrum just below threshold. Because of the
long-range character of the polarization potential, the last $s$-wave
bound state is always located relatively close to the threshold, with
the binding energy $E_b \leq 106 E^\ast = 3.02$ MHz. As one can observe
in the figure the last bound state in the open channel can be strongly
coupled to the other bound states crossing the threshold, giving rise to
relatively strong avoided crossing, as in the case of resonance E.
In contrast, very close to threshold the energy of the bound
state bends and follows the universal behavior $E = - \hbar^2(2 \mu a^2)$.

An approximate but highly accurate description of the Feshbach
resonances can be  developed using a two channel description based on CI
(configuration interaction) model of a Feshbach resonance,
where a single closed channel represents the effects of all the closed
channels contributing to a resonance \cite{Mies2000}. In the case of a single open and
a single closed channel, the application of MQDT is straightforward,
and after some simple algebra (see for instance \cite{Paul1}), one
obtain the following expression for the phase shift $\xi(E,l)$ in the
open channel \cite{JulienneFeshbachMQDT}
\begin{equation}
\label{ResPhase}
\xi(E,\ell) = \xi_\mrm{bg}(E,l) - \tan^{-1} \left( \frac{\frac\Gamma 2 C^{-2}(E,\ell)}{E-E_{n}+\frac \Gamma 2 \tan \lambda(E,\ell)}\right).
\end{equation}

The first term $\xi_\mrm{bg}(E,l)$ is the background phase shift,
describing the scattering from the open channel only and incorporating
effectively the influence of the closed channels for magnetic fields far
from the resonance. The second term describes the resonant contribution
resulting from a bound state in the closed channel with energy $E_{n}$,
crossing the threshold of the open channel. An energy-dependent width of
the resonance is given by a constant width $\Gamma$ multiplied by the
MQDT function $C^{-2}(E,\ell)$, which accounts for a proper threshold
behavior as $k\rightarrow0$. In the two-channel model one
assumes that the energy of a bound state varies approximately linearly
with the magnetic field $B$:
\begin{equation}
\label{En}
E_n(B) = \delta\mu (B -B_n),
\end{equation}
where $B_n$ is the magnetic field at which the bound state crosses the
threshold of the open channel, and $\delta\mu$ is the difference
of magnetic moments between the open and closed channel. The second
MQDT function $\tan \lambda(E)$ in the denominator of $\xi_\mrm{res}$,
describes the shift of the resonance position from the bare value $B_n$,
which is due to the coupling between the open and closed channel.

In the following we will focus on $s$-wave Feshbach resonances,
considering the zero-energy limit. According to the results of
Sections~\ref{Sec:scatt} and \ref{Sec:MQDTfun}, for $\ell =0$ the phase
shift and MQDT functions exhibit the following threshold behavior
\begin{align}
\label{Thr1}
\tan \xi(E) &\stackrel{E \rightarrow 0}{\sim} - k a - \frac{\pi}{3} (k R^\ast)^2, \\
C^{-2}(E) &\stackrel{E \rightarrow 0}{\sim} k R^\ast (1 + s^2) - (k R^\ast)^3 s^2 (1 + s^2), \\
\label{Thr3}
\tan \lambda(E) &\stackrel{E \rightarrow 0}{\sim} - s + (k R^\ast)^2 s (1+s^2),
\end{align}
where $s= a/R^\ast$. The connection to the standard theory of magnetic Feshbach resonances can be done by introducing the width of magnetic Feshbach resonance $\Delta$:
\begin{equation}
\label{DB}
\lim_{E \rightarrow 0} \frac{\Gamma}{2} \frac{C^{-2}(E)}{\tan \xi_\mrm{bg}(E)} = - \delta \mu \, \Delta,
\end{equation}
and the resonance position $B_0$, that is shifted from $B_n$ due to the coupling between the open and closed channel
\begin{equation}
\label{B0}
B_0 = B_n + \frac{\Gamma}{2 \delta \mu} \lim_{E \rightarrow 0} \tan \lambda(E).
\end{equation}
Now, substituting Eqs.~\eqref{En}, \eqref{DB}, \eqref{B0} into \eqref{ResPhase} in the limit of zero kinetic energy one retrieves the standard expression
\begin{equation}
\label{aB}
a(B) = a_\mrm{bg}\left[1-\frac\Delta{B-B_0}\right].
\end{equation}
Making use of the MQDT expansions Eqs.~\eqref{Thr1}-\eqref{Thr3} one can express $\Gamma$ and $B_n$ in terms of parameters $a_\mrm{bg}$, $\Delta$, $B_0$, $\delta \mu$ which can be directly measured in experiments
\begin{align}
B_n = B_0 + \frac{s_\mrm{bg}^2}{1+s_\mrm{bg}^2} \Delta, \qquad
\Gamma = \frac{2s_\mrm{bg}}{1+s_\mrm{bg}^2} \delta \mu \Delta.
\end{align}
Here, $s_\mrm{bg}= a_\mrm{bg}/R^\ast$.

Another important parameter characterizing Feshbach resonances is
the fraction of the closed channel in the weakly bound molecular
state at large and positive values of the scattering length
\cite{Chin2010} \begin{equation} Z(B) = \frac{1}{\delta \mu}
\frac{\partial(-E_b)}{\partial B} = \frac{1}{\zeta} \left| \frac{B_0 -
B}{\Delta}\right| \end{equation} The parameter $\zeta$ describes the
range of magnetic fields expressed as a fraction of the resonance width
$\Delta$ over which the resonance exhibit the universal properties and
the occupation of the closed channel remains small. Sufficiently close
to the resonance, the binding energy is given by the universal formula $E = - \hbar^2(2 \mu a^2)$,
which leads to \cite{Chin2010}
\begin{equation}
\label{zeta}
\zeta =
\frac{s_\mrm{bg}^2}{2} \frac{\left| \delta \mu \Delta \right|}{E^\ast} .
\end{equation}
For entrance channel dominated resonance $\zeta \gg 1$ and
$Z(B)$ remain small for detuning of the order of $\Delta$. In the opposite
case $\zeta \ll 1$, the Feshbach resonance is called closed-channel
dominated, and the universal regime where the energy-independent formula
\eqref{aB} is applicable is very narrow.

We have fitted the universal formula \eqref{aB} to Feshbach resonances
presented in Fig.~\ref{Fig:Feshbach0}. The results are summarized
in Table.~\ref{TabFesh}. The difference of the magnetic moments
has been determined from the bound state spectrum (lower panel
of Fig.~\ref{Fig:Feshbach0}) by extracting the linear slope for the
molecular states giving rise to the resonances. Finally, the parameter
$\zeta$ has been calculated from Eq.~\eqref{zeta}. We note that only
the two first Feshbach resonances, occurring at relatively small magnetic
fields, are non universal, while the remaining ones are relatively broad
and entrance-channel dominated.

\Table{\label{TabFesh} Parameters of Feshbach resonances shown in
Fig.~\ref{Fig:Feshbach0}: resonance position $B_0$, resonance width $\Delta$, background scattering length $a_\mrm{bg}$, difference of magnetic moments between the open and closed channel $\delta \mu$, and parameter $\zeta$ characterizing the fraction of the closed channel (see Eq. \eqref{zeta}). The labels in
the first column enumerate the resonances in accordance with markings in  Fig.~\ref{Fig:Feshbach0}.}
\br
 & $B_0$ (G) & $\Delta$ (G) & $a_\mrm{bg}$ ($a_0$) & $\delta \mu$ (MHz/G) & $\zeta$ \\
\mr
A & 0.322 & -0.000417 & -2019 & 4.78 & 0.0328  \\
B & 5.80 & -0.00690 & -1996 & 4.76 & 0.530 \\
C & 29.6 & -0.105 & -1919 & 4.46 & 7.00 \\
D & 91.0 & -1.38 & -1787 & 3.15 & 79.6 \\
E & 201 & -10.3 & -1803 & 1.11 & 151 \\
\br
\endTable

\subsection{Magnetic Feshbach resonances at finite energies}

We analyze here the dependence of elastic and inelastic rates on the
magnetic field at finite collision energy, where several partial waves
play in general an important role in the collision physics. To this
aim we use both our MQDT model and numerical calculations based on the
close-coupled (CC) Schroedinger equation. In order to calculate the charge
exchange rates from the CC solutions we have generalized the approximate
closure relation Eq.~(\ref{closure}) to the multichannel case, including
hyperfine effects and the external magnetic field. The charge-exchange
process in the MQDT model has been described using the semiclassical
approximation Eq.~\eqref{SigmaTr2} where the MQDT function $C^{-2}(E)$
is replaced by the amplitude of the wave function in the singlet $A^1
\Sigma$ channel. Details are provided in Appendix~\ref{App:Multi}.

Figs.~\ref{Fig:Feshbach1nK}, \ref{Fig:Feshbach1muK} and
\ref{Fig:Feshbach1mK} show the elastic and charge transfer rates at three
collision energies: $E = 1$nK, $E=1 \mu$K and $E=1$mK for some typical
values of the singlet and triplet scattering length, $a_s =R^\ast=2081
a_0$ and $a_t =-R^\ast$.  Contributions from the lowest 2, 9 and 21
partial waves are included respectively at these three energies. At 1nK
resonances appear mostly in the $s$-wave and are relatively broad. In
addition one can observe few narrow resonances in the $p$-wave channel
occurring at $B = 27$, $84.7$ and $182$G. In this range of temperatures
the MQDT model is extremely accurate and agrees perfectly with the full
numerical CC calculations.

At higher energy $E=1\mu$K the agreement of the analytic model with the
numerical solution is still very good. At this energy the resonance peaks
arise from the 4 lowest partial waves. The arrow in the bottom panel
of Fig.~\ref{Fig:Feshbach1muK} indicates the $\ell = 3$ Feshbach
resonance. It is the only resonance at $E=1\mu$K whose position is not
well predicted by the analytical MQDT model. We note that the charge
transfer rates exhibit more resonance peaks than the elastic rates. This
general behavior can be qualitatively understood by analyzing the number
of partial waves contributing to the elastic and inelastic processes. The
elastic rates are dominated by the reflection from the long-range
polarization potential and their contribution to the cross section
decay as $(2 \ell+1) \sin^2 \xi_\ell (k) \sim 1/\ell^5$ at large $\ell$
(see Eq.~\eqref{xil1}). In contrast the charge exchange process must
involve the tunneling through the centrifugal barrier, and its probability
decays exponentially with $\ell$. Hence the number of partial waves
contributing to the inelastic process is much smaller than for elastic
scattering, and the inelastic rates are more sensitive to scattering
resonances. On the other hand, the narrow resonances from high-order partial
waves that appear in the elastic rates are less pronounced due to the
strong background arising from reflection on the long-range potential.

Finally, at the highest energy considered $E=1$mK the resonances are
narrower and have smaller amplitude. This effect arises
due to the large number of partial waves contributing to the scattering,
which have tendency to wash out the resonance structures. We observe that
at 1mK the MQDT model basically follows the magnetic field dependence of
the exact numerical rates, but it predicts accurately only the resonances
associated to the lowest partial waves. We note that inclusion of the
thermal averaging washes out the resonance structure for the elastic
rates, while the charge-transfer rates still exhibit some resonance
peaks. This is again due to the vastly different numbers of partial
waves contributing to the elastic and inelastic collision processes.

\begin{figure}
\begin{indented}
\item[]\includegraphics[width=8.6cm,clip]{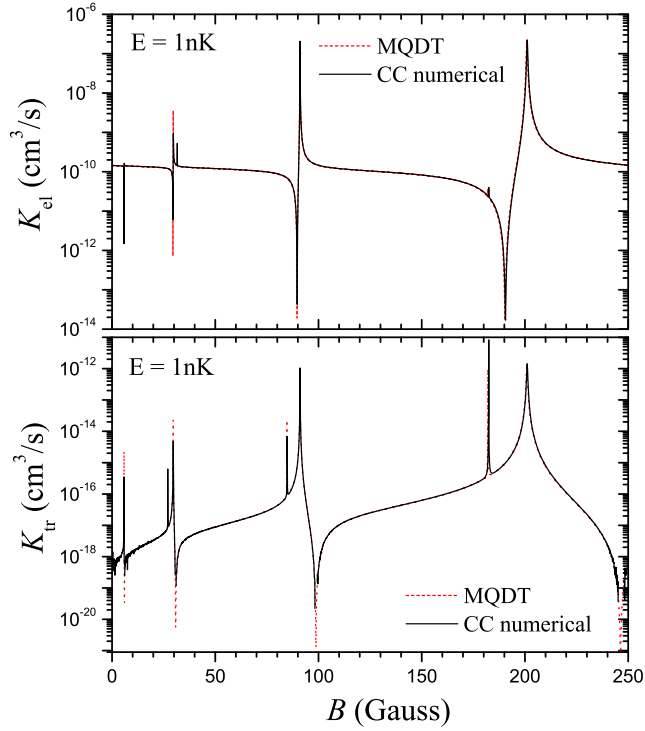}
\end{indented}
\caption{
\label{Fig:Feshbach1nK}
Elastic (upper panel) and charge exchange (lower panel) collision rates
for collisions of Na and $^40$Ca$^+$ versus magnetic field calculated
at energy $E=1$nK for singlet and triplet scattering lengths $a_s
=R^\ast=2081 a_0$ and $a_t =-R^\ast = -2081 a_0$. The figure compares
the CC numerical calculations (black solid) and the quantum-defect model
(red dashed).}
\end{figure}
\begin{figure}
\begin{indented}
\item[]\includegraphics[width=8.6cm,clip]{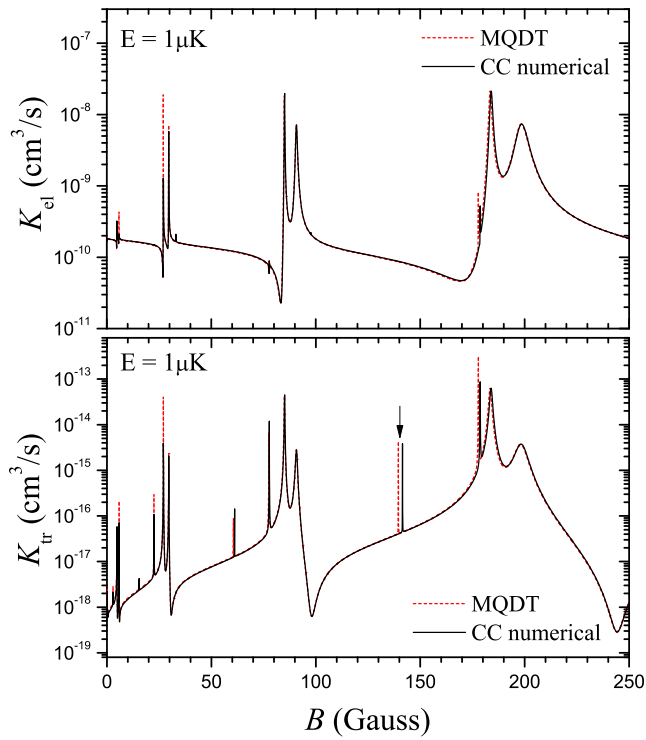}
\end{indented}
\caption{
\label{Fig:Feshbach1muK}
Same as Fig.~\ref{Fig:Feshbach1nK} but for collision energy
$E=1\mu$K. The arrow in the lower panel indicates the position of the
$\ell=3$ Feshbach resonance (see text for details).}
\end{figure}
\begin{figure}
\begin{indented}
\item[]\includegraphics[width=8.6cm,clip]{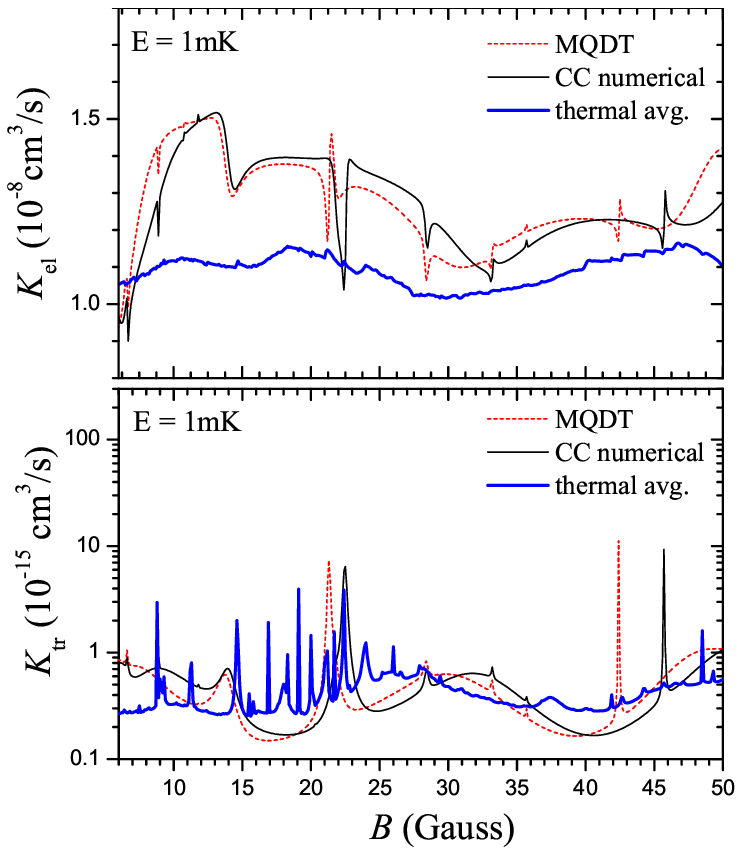}
\end{indented}
\caption{
\label{Fig:Feshbach1mK}
Same as Fig.~\ref{Fig:Feshbach1nK} but for collision energy $E=1$mK. The thick blue line represents the MQDT result averaged over a thermal distribution with temperature of 1mK.}
\end{figure}

\section{Conclusions}

Summarizing, we have developed a quantum-defect model for ultracold
atom-ion collisions. The model was applied to the reference system composed by a
${}^{40}\mrm{Ca}^{+}$ ion and a ${}^{23}\mrm{Na}$ atom, and its predictions
were thoroughly verified by comparison with numerical close-coupled
calculations using {\it ab initio} potential energy curves. Our model is based
on the multichannel quantum-defect formalism, where the quantum-defect
parameters are defined in terms of the analytic solutions for $r^{-4}$
polarization potentials.  Use of a frame-transformation allows us to reduce
the number of short-range parameters to essentially singlet and triplet
scattering lengths only. Since for atom-ion systems of experimental
interest the values of the singlet $a_s$ and triplet $a_t$
scattering lengths are not yet known, in our calculations we have assumed
typical scattering lengths of the order of the characteristic length
$R^\ast$. Once $a_s$ and $a_t$ will be measured our model could be
readily applied to obtain all the basic collisional properties in the
ultracold domain, including accurate positions of Feshbach and shape
resonances. Application of our theory to other atom-ion systems is
straightforward, amounting to a simple change of the scattering lengths
and to the use of new characteristic parameters $R^\ast$ and $E^\ast$, which
are determined by the atomic polarizability and the reduced mass.

In our studies we considered only two-body collisions in free space
ignoring possible effects of the trapping potential, which is present in
realistic systems. This can be of particular importance for the ions, that
acquire in the presence of a time-dependent radio-frequency potential
a small amplitude high frequency motion known as micromotion. This
effect is important for current experiments, leading on one side to a
significant loss of atoms, on the other potentially preventing sympathetic
cooling of the ions to the ground-state of the trapping potential
\cite{DeVoe2010}. Moreover, the presence of a tight ion trap
with characteristic size smaller than $R^\ast$ modifies the long-range
asymptotics of the atom-ion wave functions and results in principle in
the coupling of relative and center-of-mass motion. In addition, in the
presence of a magnetic field the charged ions describe cyclotron orbits,
an additional confinement effect which can lead to the appearance of scattering
resonances \cite{Andrea2010} and may affect the actual positions of
Feshbach resonances. We are currently investigating these issues.

\ack
The authors thank J. Denschlag and C. Sias for helpful discussions. This work was supported by CNRS (Z.I. and A.S.), Rennes Metropole (A.S.), the Polish Government Research Grant for years 2007-2010 (Z.I.), and the European project AQUTE (T.C.).

\appendix

\section{Properties of analytical solutions for $r^{-4}$ potential}
\label{App:Mathieu}

\subsection{Basic derivation}

We start from the Mathieu's equation of the imaginary argument \eqref{Mathieu}
\begin{equation}
\label{Mathieu2}
\frac{d^2 \psi}{d z^2} - \left[ a - 2 q \cosh 2 z \right] \psi = 0.
\end{equation}
where $a=(l+{\tst \frac 12})^2$ and $q=\sqrt{E}$. It is convenient to look for the solution
of \eqref{Mathieu} in the following form
\begin{equation}
\label{DefM}
M_\nu(z) =  \sum_{n=-\infty}^{\infty} c_n(\nu) e^{(2n+\nu)z}, \\
\end{equation}
where $\nu$ is the characteristic exponent. Substituting the ansatz \eqref{DefM} into \eqref{Mathieu} we obtain the following
recurrence relation
\begin{equation}
\label{rec}
\left[(2n+\nu)^2 - a \right] c_n + q (c_{n-1} + c_{n+1}) = 0.
\end{equation}
The three-term recurrence can be solved in terms of the continued fractions. In analogy to the solution of $r^{-6}$ potentials \cite{Gao} we subsitute
\begin{align}
\label{defbp}
c_{n} & = \left(-\frac{q}{4}\right)^n \frac{\Gamma\left(\frac{\nu - \sqrt{a}}{2}+1\right)
\Gamma\left(\frac{\nu+\sqrt{a}}{2}+1\right)}
{\Gamma\left(\frac{\nu-\sqrt{a}}{2}+1+n\right) \Gamma\left(\frac{\nu+\sqrt{a}}{2}+1+n\right)} b_n^{+} \\
\label{defbm}
c_{-n} & = \left(-\frac{q}{4}\right)^n \frac{\Gamma\left(\frac{\nu-\sqrt{a}}{2}-n\right)
\Gamma\left(\frac{\nu+\sqrt{a}}{2}-n\right)}
{\Gamma\left(\frac{\nu-\sqrt{a}}{2}\right)
\Gamma\left(\frac{\nu+\sqrt{a}}{2}\right)} b_n^{-}
\end{align}
for $n\geq 0$. In the case of $n=0$ we have $c_0 =  b_0^{+} =  b_0^{-}$. Now the recurrence relation \eqref{rec} can be
written as
\begin{align}
b_n^{+} - b_{n-1}^{+}= \frac{q^2 b_{n+1}^{+}}{\left[(2n+2+\nu)^2-a \right] \left[(2n+\nu)^2-a \right]}, \\
b_n^{-} - b_{n-1}^{-}=  \frac{q^2 b_{n+1}^{-} }{\left[(2n+2-\nu)^2-a \right] \left[(2n-\nu)^2-a \right]},
\end{align}
Finally we substitute $h_n^{+} = b_n^{+} / b_{n-1}^{+}$ and  $h_n^{-} = b_n^{-} / b_{n-1}^{-}$, which yields the continued
fractions
\begin{align}
\label{rec1}
h_n^{+} & = \frac{1}{1-\frac{q^2}{\left[(2n+2+\nu)^2-a \right] \left[(2n+\nu)^2-a \right]}  h_{n+1}^{+}}, \\
\label{rec2}
h_n^{-} & = \frac{1}{1-\frac{q^2}{\left[(2n+2-\nu)^2-a \right] \left[(2n-\nu)^2-a \right]}  h_{n+1}^{-}}
\end{align}
To find values of the coefficients $c_n$ it is sufficient to set $h^{+}_{m}=1$ and $h^{-}_{m}=1$ for some,
sufficiently large $m$ and calculate $h^{+}_{n}$ and $h^{-}_{n}$ up to $n=1$ using \eqref{rec1} and \eqref{rec2}. Then
\begin{align}
\label{bp}
b_n^{+} & = h_n^{+}  h_{n-1}^{+} \ldots  h_{1}^{+} c_0, \\
\label{bm}
b_n^{-} & = h_n^{-}  h_{n-1}^{-} \ldots  h_{1}^{-} c_0,
\end{align}
and coefficients $c_n$ can be obtained from Eqs.~\eqref{defbp}-\eqref{defbm}.
Characteristic exponent $\nu$ has to determined from Eq.~\eqref{rec} with $n=0$:
\begin{equation}
\label{nu1}
\nu^2 - a  - q^2 \left(\frac{h_{1}^{+}(\nu)}{(\nu+2)^2-a}  + \frac{h_{1}^{-}(\nu)}{(\nu-2)^2-a} \right) = 0,
\end{equation}

In numerical calculations it is more convenient to find $\nu$ from equation \cite{Erdelyi}
\begin{equation}
\cos \pi \nu = 1 - \Delta (1 - \cos \pi \sqrt{a} )
\end{equation}
where $\Delta$ is an infinite determinant (independent of $\nu$)
\begin{equation}
\Delta =
\left| \begin{array}{ccccccc}
\ddots & \vdots & \vdots & \vdots & \vdots & \vdots &  \\
\ldots & 1 & \gamma_{-2} & 0 & 0 & 0 & \ldots \\
\ldots & \gamma_{-1}& 1 & \gamma_{-1} & 0 & 0 &\ldots \\
\ldots & 0 & \gamma_0 & 1 & \gamma_0 & 0 & \ldots \\
\ldots & 0 & 0 & \gamma_1 & 1 & \gamma_1 & \ldots \\
\ldots & 0 & 0 & 0 & \gamma_2 & 1  & \ldots \\
 & \vdots & \vdots & \vdots & \vdots & \vdots & \ddots
\end{array} \right|
\end{equation}
with $\gamma_n= q/(4n^2 - a)$.
Typically determinant $\Delta$ converges very fast and to calculate $\Delta$ it is enough to take relatively small matrices.

\subsection{Asymptotic expansions for large arguments}

To derive asymptotic expansion of $M_{\nu}(z)$ for $z \rightarrow \infty$ we observe that the leading contribution to
the sum \eqref{DefM} comes from the terms with large $n$. We neglect contribution of terms with $n<0$ and
apply the following approximation for the terms with $n\geq 0$
\begin{equation}
\label{cpapprox}
c_{n} \approx \left(-\frac{q}{4}\right)^n \frac{\Gamma\left(\frac{\nu - \sqrt{a}}{2}+1\right)
\Gamma\left(\frac{\nu+\sqrt{a}}{2}+1\right)}
{\Gamma\left(\frac{\nu-\sqrt{a}}{2}+1+n\right) \Gamma\left(\frac{\nu+\sqrt{a}}{2}+1+n\right)} b_{\infty}^{+} \\
\end{equation}
where $b_{\infty}^{+} = \lim_{n \rightarrow \infty} b_{n}^{+}$. This yields
\begin{align}
M_\nu(z) \stackrel{z \rightarrow \infty}{\longrightarrow} & \,
b_{\infty}^{+} \,
\Gamma\left(\frac{\nu-\sqrt{a}}{2}+1\right) \Gamma\left(\frac{\nu+\sqrt{a}}{2}+1\right) \left( \frac{2}{\sqrt{q}} \right)^{\nu} J_{\nu}(\sqrt{q} e^z),
\label{Mazp}
\end{align}
where $ J_{\nu}$ denotes Bessel function. Now the asymptotic behavior for large $z$ can be easily obtained from the well-known
asymptotic expansions of the Bessel funtions.

In similar way we obtain behavior for large and negative $z$
\begin{align}
M_\nu(z) \stackrel{z \rightarrow - \infty}{\longrightarrow} & \,
b_{\infty}^{-} \,
\Gamma\left(1-\frac{\nu-\sqrt{a}}{2}\right) \Gamma\left(1-\frac{\nu+\sqrt{a}}{2}\right) \left( \frac{\sqrt{q}}{2} \right)^{\nu} J_{-\nu}(\sqrt{q} e^{-z}),
\label{Mazn}
\end{align}
where $b_{\infty}^{-} = \lim_{n \rightarrow \infty} b_{n}^{-}$.

\subsection{Two linearly independent solutions}
\label{App:TwoSol}

As the two linearly independent solutions of \eqref{Mathieu} we can take $M_{\nu}(z)$ and $M_{-\nu}(z)$
\footnote{It is easy to oberve from Eq.~\eqref{rec} that if $\nu$ is a characteristic exponent then $-\nu$ must be also
a characteristic exponent and $c_n(-\nu) = c_{-n}(\nu)$}. It is convenient for our purposes to define the following two linearly
independent solutions in initial variable $r$
\begin{align}
T_\nu(r) & = w(\nu) M_{\nu} \left[ \ln \left(\sqrt{q} r\right) \right] \sqrt{r}, \\
T_{-\nu}(r) & = w(-\nu) M_{-\nu} \left[ \ln \left(\sqrt{q} r\right) \right] \sqrt{r},
\end{align}
where
\begin{equation}
w(\nu) =  \sqrt{\frac{\pi}{2}} \frac{(4/q)^{\nu/2}}{b_{\infty}^{+}(-\nu)
\Gamma\left(\frac{l-\nu}{2}+\frac54\right) \Gamma\left(\frac34-\frac{\nu+l}{2}\right) }
\end{equation}

Using Eqs.~\eqref{Mazp} and \eqref{Mazn} one can easily work out
asymptotic behavior of $T_{\nu}(r)$. For small $r$ we obtain
\begin{align}
\label{Rprs}
T_{\nu}(r) \stackrel{r \rightarrow 0}{\longrightarrow} &
r \cos \left( \frac{1}{r} + \frac \pi 2 \nu  - \frac \pi 4 \right),
\end{align}
whereas for large $r$ we get
\begin{align}
\label{Rprl}
T_{\nu}(r) \stackrel{r \rightarrow \infty}{\longrightarrow} & S(\nu)  \left( \frac{4}{q} \right)^{\nu} \frac{
\cos \left( k r - \frac \pi 2 \nu  - \frac \pi 4 \right)}{\sqrt{q}},
\end{align}
where
\begin{equation}
\label{DefS}
S(\nu)=  \frac{b_{\infty}^{+}(\nu)}{b_{\infty}^{-}(\nu)}\frac{\Gamma\left(\frac{\nu+l}{2}+\frac54\right)
\Gamma\left(\frac{\nu-l}{2}+\frac34\right)}
{\Gamma\left(\frac{l-\nu}{2}+\frac54\right) \Gamma\left(\frac34 -\frac{\nu+l}{2}\right)}
\end{equation}

\subsection{Expansions for small q}
\label{App:ExpSmallQ}

The small-$q$ expansion of the characteristic exponent $\nu$ can be obtained from Eq.~\eqref{nu1}
\begin{equation}
\label{nuex}
\nu =  l+ \frac12 - \frac{q^2}{4(l-\frac12)(l+\frac12)(l+\frac32)} + O(q^4).
\end{equation}
Utilizing this results and applying \eqref{rec1}-\eqref{rec2} and
\eqref{bp}-\eqref{bm} we calculate expansions of $b_{\infty}^{+}(\nu)$ and $b_{\infty}^{-}(\nu)$
\begin{align}
b_{\infty}^{+}(\nu) = & 1 + \frac{ (l + \frac72)(l+\frac12)-(2l+3)[\gamma+\psi(l+\frac32)]}{
(2l+3)^2(2l-1)(2l+1)} q^2 
\label{bpex} +  {\cal O}(q^4), \\
b_{\infty}^{-}(\nu) = & 1 +  \frac{ (l+\frac12)(l-\frac52)+(2l-1)[\gamma+\psi(\frac12-l)]}{
(2l-1)^2(2l+3)(2l+1)} q^2 
\label{bmex} +  {\cal O}(q^4),
\end{align}
where $\psi(x)$ denotes the digamma function.

Finally expansion of $S(\nu)$ can be obtained from definition \eqref{DefS} where we substitute \eqref{bpex}, \eqref{bmex}
and expand the Gamma functions. This yields
\begin{equation}
S(\nu) = \frac{\Gamma(\frac32+l)}{\Gamma(\frac12-l)}\Big[ 1 -
\frac{(l-\frac12)(l+\frac32)(\psi(\frac12-l) + \psi(l+\frac32))-(l+\frac12)^2}{
4 (l-\frac12)^2 (l+\frac12) (l+\frac32)^2} q^2 +  O(q^4) \Big].
\label{sex}
\end{equation}

\section{Multichannel calculations of the radiative charge transfer}
\label{App:Multi}

Equation~(\ref{closure}) based on the approximated closure relation can be
easily generalized it to the multichannel case including hyperfine effects
and an external magnetic field. In this case the initial state for atoms
incoming in the dressed channel state $\alpha$ of Eq.~\eqref{BDressStates}
will be labeled $|E \ell \alpha \rangle$. The reduced dipole moment
becomes a matrix, with only nonvanishing diagonal elements if the total
electron and nuclear spin $\{S I \}$ representation is used.

In the presence of a magnetic field close-coupled scattering equations
are numerically solved with a matrix perturbation $\xi \bar{A}(r)$.
The emission rate for atoms incoming at collision energy $E$ is expressed
in the DWBA in terms of the open-open scattering matrix elements $s_{\xi
\beta \alpha}=s (E \ell  \beta \leftarrow E \ell  \alpha; \xi) $ with
and without field perturbation
\be
\sum_\beta {s^*_{0\alpha \beta}}(s_{\xi \beta \alpha} -
 s_{0 \beta \alpha})
= -2\pi i \xi  \langle E \ell \alpha |  \bar{A}(r) | E \ell \alpha \rangle  .
\ee
As in the single-channel case, right multiplication by the ${\bf s}^*_0$
matrix enforces the correct boundary conditions, and symmetry of the
scattering matrix ($s_{\xi \alpha \beta}=s_{\xi \beta \alpha}$) resulting
from time-reversal invariance has been used.

Calculation of the charge-exchange rates in the MQDT approach can be done
using the semiclassical formula Eq.~\eqref{SigmaTr2}, with the function
$C^{-1}(E,\ell)$ replaced by the amplitude $A_s^{(IS)}(E,\ell)$ of the
singlet component of the multichannel wave function at short range
\begin{equation}
\label{SigmaTr3}
\sigma_\mrm{tr}(E) = \frac{2 \pi}{k^2} P_\mrm{tr} \sum_{\ell} (2 \ell + 1) |A_s^{(IS)}(E,\ell)|^2
\end{equation}
In order to calculate $A_s^{(IS)}$ we first analyze the open-open block of the multichannel wave function at large distances
\begin{equation}
\label{Foo1}
\mbf{F}_\mrm{oo}(r) \stackrel{r \rightarrow \infty}{\longrightarrow} \left[\hat{\mbf{f}}_\mrm{oo}(r) + \hat{\mbf{g}}_\mrm{oo}(r) \hat{\mbf{Y}}_\mrm{oo} \right]
\mbf{A}_\mrm{oo}
\end{equation}
Using relations between short-range and long-range normalized solutions, Eq.~\eqref{Foo1} can be rewritten as
\begin{equation}
\begin{split}
\mbf{F}_\mrm{oo}(r) & \stackrel{r \rightarrow \infty}{\longrightarrow} \left[\mbf{f}_\mrm{oo}(r) + \mbf{g}_\mrm{oo}(r) \mbf{R}(E)\right]
\mbf{C}(E) \left[\mbf{1}- \tan {\bm \lambda}(E) \bar{\mbf{Y}}_\mrm{oo} \right]
\mbf{A}_\mrm{oo}
\end{split}
\end{equation}
with $\mbf{R}(E)$ given by formula~\eqref{R} applied for
the open-open block of the renormalized quantum-defect matrix
$\bar{\mbf{Y}}_\mrm{oo}$. The constant matrix $\mbf{A}_\mrm{oo}$ is fixed
by the boundary conditions at $r \to \infty$. With the following choice
of $\mbf{A}_\mrm{oo}$,
\begin{equation}
\mbf{A}_\mrm{oo}= \left[\mbf{1}- \tan {\bm \lambda}(E) \bar{\mbf{Y}}_\mrm{oo} \right]^{-1} \mbf{C}(E)^{-1} \left[\mbf{1}- i \mbf{R}(E) \right]^{-1} e^{i {\bm \xi}}
\end{equation}
the wave function has a normalization corresponding to a unit flux of incoming particles
\begin{equation}
\mbf{F}_\mrm{oo}(r) \stackrel{r \rightarrow \infty}{\longrightarrow}
\frac{1}{2} \left[
\mbf{H}^{(2)}(k r) + \mbf{H}^{(1)}(k r) \mbf{S}
\right].
\end{equation}
Here, $H^{(2)}_{ij}(kr) \rightarrow \delta_{ij} i
e^{-i(k_i r - \ell_i \pi/2)}/\sqrt{k_i}$ and
$H^{(1)}_{ij}(kr) = \left[H^{(2)}_{ij}(kr)\right]^\ast$ are functions exhibiting asymptotic behavior associated with the spherical Hankel functions $h_\ell^{(2)}(kr)$ and $h_\ell^{(1)}(kr)$, respectively.

The total wave function expressed in terms of the short-range normalized solutions reads
\begin{equation}
\label{Foo2}
\mbf{F}(r) = \left[\mbf{f}(r) + \hat{\mbf{g}}(r) \hat{\mbf{Y}} \right]
\mbf{A}
\end{equation}
with
\begin{equation}
\label{Aoc}
\mbf{A} =
\left(
\begin{array}{c}
\mbf{A}_\mrm{oo}\\
-\left[\mbf{Y}_\mrm{oo} + \tan {\bm \nu}(E) \right] \mbf{Y}_\mrm{co} \mbf{A}_\mrm{oo}
\end{array}
\right)
\end{equation}
chosen in such a way that closed channel wave function at large distance
is proportional to the exponentially decaying solution $\phi_i(r)$,
Eq.~\eqref{phi}. Applying the frame transformation yields the multichannel
amplitude of the wave function in the molecular basis
\begin{equation}
\mbf{A}^{(IS)}= \left(\mbf{Z}(B) \mbf{U} \cos {\bm \chi} \right)^{-1} \mbf{A}
\end{equation}
where ${\bm \chi}_{\beta \beta'} = \delta_{\beta \beta'} \phi_{S(\beta)}$ is the diagonal matrix containing short-range phases of the singlet $\phi_0$
and the triplet $\phi_1$ potentials, with $S(\beta)$ denoting the total electron spin in the channel $\beta$.

\section*{References}
\bibliography{mqdt}

\end{document}